\documentclass[fleqn,usenatbib]{mnras}
\usepackage[english]{babel} 
\usepackage{natbib}
\usepackage{enumerate}
\usepackage{color}

\usepackage{aas_macros}
\usepackage{epsfig}
\usepackage{pifont}
\usepackage{verbatim, amsmath, amsfonts, amssymb,amssymb}
\usepackage[utf8x]{inputenc}
\usepackage[usenames,dvipsnames,table]{xcolor}
\usepackage{booktabs, longtable}
\usepackage{multirow}
\usepackage{float}
\usepackage{lscape}
\usepackage{graphicx, subfigure}
\usepackage{rotating}
\usepackage{enumerate}
\usepackage{url}
\usepackage{bm}
\usepackage{tikz}
\usepackage{color}
\usepackage[lined]{algorithm2e}
\usepackage{listings}
\usepackage{soul}
\usepackage{tikz}
\usepackage{empheq}
\usepackage{graphicx}
\usepackage{mathrsfs}
\usepackage{times}
\usepackage{lscape}
\usepackage{subfigure}
\usepackage{lineno}
\usepackage{adjustbox}
\usepackage{breqn}

\newcommand{\vect}[1]{\boldsymbol{#1}}

\usepackage[table]{xcolor}
\usepackage{booktabs}

\def\apj{ApJ}

\newcommand{\gt}{\symbol{"3E}}	   		
\newcommand{\Tau}{\mathrm{T}}
\title{A case study of hurdle and generalized additive models in astronomy:  the escape of ionizing radiation}

\author{{\parbox{\textwidth}{\vspace{-.5cm} \Large M.~W.~Hattab$^{1}$, R.S. de Souza$^{2}$, B. Ciardi $^{3}$,  J.-P. Paardekooper$^{4}$, S. Khochfar$^{5}$, C. Dalla Vecchia$^{6,7}$     
 }}\vspace{0.3cm}\\
\parbox{\textwidth}{ \small 
 $^{1}$Wyss Institute for Biologically Inspired Engineering, Harvard Medical School, Boston, MA 02115, USA\\
 $^{2}$Department of Physics \& Astronomy, University of North Carolina at Chapel Hill, Chapel Hill, NC 27599-3255, USA\\
 $^{3}$Max-Planck-Institut f\"ur Astrophysik, Karl-Schwarzschild-Str. 1, 85748 Garching, Germany\\
 $^{4}$Universität Heidelberg, Zentrum für Astronomie, Institut für Theoretische Astrophysik, Albert-Ueberle-Str. 2, 69120 Heidelberg, Germany\\ 
 $^{5}$Institute for Astronomy, University of Edinburgh, Royal Observatory, Blackford Hill, Edinburgh EH9 3HJ, UK\\
 $^{6}$Instituto de Astrofis\'{i}ca de Canarias, C/ V\'{i}a L\'{a}ctea s/n,38205 La Laguna, Tenerife, Spain\\
 $^{7}$Departamento de Astrofis\'{i}ca, Universidad de La Laguna, Av. del Astrof\'{i}sico Franciso S\'{a}nchez s/n, 38206 La Laguna, Tenerife, Spain
 }}

\begin{document}
\maketitle
\label{firstpage}


\topmargin -1.3cm


\begin{abstract}
The dark ages of the Universe  end with the formation of the first generation of stars residing in primeval galaxies. These objects were the first to produce ultraviolet ionizing photons in a period when the cosmic gas changed from a neutral state to an ionized one,  known as Epoch of Reionization (EoR).
A pivotal aspect to comprehend the EoR is to probe the intertwined relationship between the  fraction of ionizing photons capable to escape  dark haloes, also known as the escape fraction ($f_{esc}$),  and the physical properties of the galaxy. 
This work develops a sound statistical  model suitable to  account for such non-linear relationships and the non-Gaussian nature of $f_{esc}$. This model simultaneously estimates the probability that a given primordial galaxy starts the ionizing photon production and estimates the mean level of the $f_{esc}$ once it is triggered. The model was employed in the First Billion Years simulation suite, from which
we show that the baryonic fraction and the rate of ionizing photons appear to have a larger impact on $f_{esc}$ than previously thought.  A naive univariate analysis of the same problem would suggest smaller effects for these properties and a much larger impact for the specific star formation rate, which is lessened after accounting for other galaxy properties and non-linearities in the statistical model. A snippet code to reproduce the analysis of this paper is available at \href{https://github.com/COINtoolbox/Hurdle\_fEsc}{COIN toolbox}

\end{abstract}



\section{Introduction}
\label{sec:intro}
%

The  Epoch of Reionization (EoR) represents a milestone in the evolution of the Universe marking the transition from  an initially neutral inter-galactic medium (IGM) into a hot, ionized plasma \citep[e.g.][]{2000ApJ...535..530G,2003MNRAS.343.1101C,2014MNRAS.439..725I,2018MNRAS.476.1174E,Dayal2018}. The early generations of stars \citep{Schneider2002,Maio2010f,2013MNRAS.436.1555D,Schauer2015} are likely among the major players of this transformation, which thus strongly depends on the (poorly known) high-$z$ stellar properties. Among these, the escape fraction of ionizing photons, $f_{esc}$\footnote{The escape fraction is defined as the fraction of H-ionizing (i.e. with energy above 13.6~eV) photons emitted within a galaxy that is able to escape into the intergalactic medium.}, 
represents a key parameter to probe different  reionization scenarios in observational \citep[e.g.][]{2012ApJ...752L...5B,2013ApJ...768...71R}, semi-analytical  \citep[e.g.][]{Choudhury:2009db,Pritchard:2010br,Santos:2010hr,Mesinger:2011gm,2012ApJ...747..100S,2013MNRAS.428L...1M} as well as simulated  \citep[e.g.][]{Iliev2006,Trac2007,2012MNRAS.423..558C} investigations. Despite its relevance, $f_{esc}$ remains a very uncertain quantity.  
Current observations indicate a level of  $f_{esc}$ lower than the one required for galaxies to contribute significantly to the EoR (but see e.g. \citealt{2018MNRAS.476L..15V}) suggesting that either other sources of ionizing radiation contribute significantly, or low redshift galaxies are not representative of the galaxies from the EoR.

Primordial galaxies are mostly below the observational detection limit at redshifts $z \gt 6$, therefore an investigation of $f_{esc}$ during the early Universe necessarily relies on theoretical models. Previous numerical simulations have found evidence for a strong dependence of $f_{esc}$ on galaxy mass and redshift  \citep{2000ApJ...545...86W,2000ApJ...542..548R,2003ApJ...599...50F,Yajima2011,2013MNRAS.429L..94P,Wise2014,Paardekooper2015}.  Since massive galaxies offer a higher column density to ionizing photons, $f_{esc}$ is expected to decrease with increasing mass of the host galaxy. On the other hand, their deeper potential wells may trigger a higher star formation, leading to an increment in $f_{esc}$  \citep{Gnedin2008,Shull2012,Benson2013}. Such complex relationships hold for other galaxy properties \citep{Paardekooper2015}, but no clear dependence was found \citep{Yajima2014}. This motivates the development of a  statistical approach capable to characterize the dependence of $f_{esc}$ on various galaxy physical properties without relying upon oversimplifying approximations. 

This work introduces a hurdle Binomial-Beta generalized additive model to approximately accommodate the complex relationships between $f_{esc}$ and halo properties, while respecting the fractional nature of $f_{esc}$, not amenable by Gaussian (normal) approximations. 
Our aim, though, is not only to introduce a novel approach to probe $f_{esc}$, but also to show how linear models can be easily extended to accommodate various types of response variables and non-linear relationships by adding layer by layer of complexity. In fact, while linear models are ubiquitous in astronomical research, their extensions (including mathematical details, fitting and inferential procedures) are yet to be fully exploited.
Most of the discussion throughout this paper is therefore new in the astronomical literature, but yet widely applicable to different kinds of inferential  problems \citep[see e.g.][for a general introduction of generalized linear models and extensions for astronomers]{Hilbe2017}.

The outline of the paper is as follows. 
Section \ref{sec:catalog} describes the dataset and the variables of interest that are sought for to approximate $f_{esc}$. Section~\ref{sec:bbgamh} briefly introduces the formalism of linear models, its limitations, and the motivation behind the use of generalized linear models, logistic regression, beta regression, hurdle models and finally generalized additive models. A hurdle Binomial-Beta generalized additive model is then presented. This model is fitted to the $f_{esc}$ data, and the results are shown in section~\ref{sec:Results} along with a statistical method to rank the predictors according to their importance to define the behaviour of $f_{esc}$. Discussion and conclusions are given in Section~\ref{sec:results_discussion}.

\section{Dataset}
\label{sec:catalog}

The dataset used in this work is retrieved from  the First Billion Years (FiBY) simulation suite \citep[first described in][]{2013MNRAS.429L..94P}, and is based on the catalogue built by  \citet{Paardekooper2015}.
The FiBY simulation suite used a customized version of the smoothed-particle hydrodynamics code {\sc gadget} \citep{Springel2005} tailored for the {\it Overwhelmingly Large Simulations} project \citep{Schaye2010}.  Haloes were identified with a  {\sc subfind} algorithm \citep[e.g.][]{2009MNRAS.399..497D}, and their redshift evolution was followed down to $z=6$ using a merger tree approach \citep{2012MNRAS.421.3579N}.  The gas cooling was evaluated using tables for line-cooling in photo-ionization equilibrium computed for 11 elements: H, He, C, N, O, Ne, Mg, Si, S, Ca and Fe \citep{2009MNRAS.393...99W},  with the {\sc cloudy} v07.02 code \citep{2000RMxAC...9..153F}. 
 The  FiBY simulations included an additional full non-equilibrium primordial chemistry network \citep{Abel1997,Yoshida2006,Maio2011MNRAS.414,Johnson2013MNRAS} with molecular cooling functions for H$_2$ and HD, a thermal SN II feedback model \citep{Dalla2012}, and a time-step limiter algorithm  to preserve the concordance of feedback methods \citep{Durier2012,Schaller2015MNRAS.454}.  The FiBY simulations have been shown to reproduce general properties of the galaxy population such has the star formation rate main sequence and mass function well  \citep[see e.g.][]{Cullen2017,Agarwal2018}.
The simulated catalogue combines the FiBY\_S simulation and the FiBY simulation. FiBY\_S has a box size of $4$ Mpc and is composed of $2 \times 684^3$ dark matter and gas particles. It was run until redshift $6$. The FiBY simulation has a box size of $8$ Mpc, contains $2 \times 1368^3$ dark matter and gas particles,  and was run until redshift 8.5. 
 In every halo the escape fraction was computed by comparing the number of photons that are produced by the stars,  $N_{\mathrm{emitted}}$, to the number of photons that reach $r_{200}$ of the main halo,  $N_{\mathrm{phot}}(r > r_{200})$:
 \begin{equation}\label{eq:fEsc}
   f_{\mathrm{esc}} = \frac{N_{\mathrm{phot}}(r \ge r_{200}) }{N_{\mathrm{emitted}}},
 \end{equation}
 in which $r_{200}$ is the radius at which the overdensity is 200 times the critical density.
A  $\Lambda$CDM cosmology: $\Omega_{\textnormal{M}} = 0.265$, $\Omega_{\Lambda} = 0.735$, $\Omega_{\textnormal{b}} = 0.0448$, $H_0 = 71 \rm{\,km} \rm{\ s}^{-1} \rm{Mpc}^{-1}$, and  $\sigma_8 = 0.81$ \citep{2009ApJS..180..330K}, has been used.

\begin{figure}
\includegraphics[width=\linewidth]{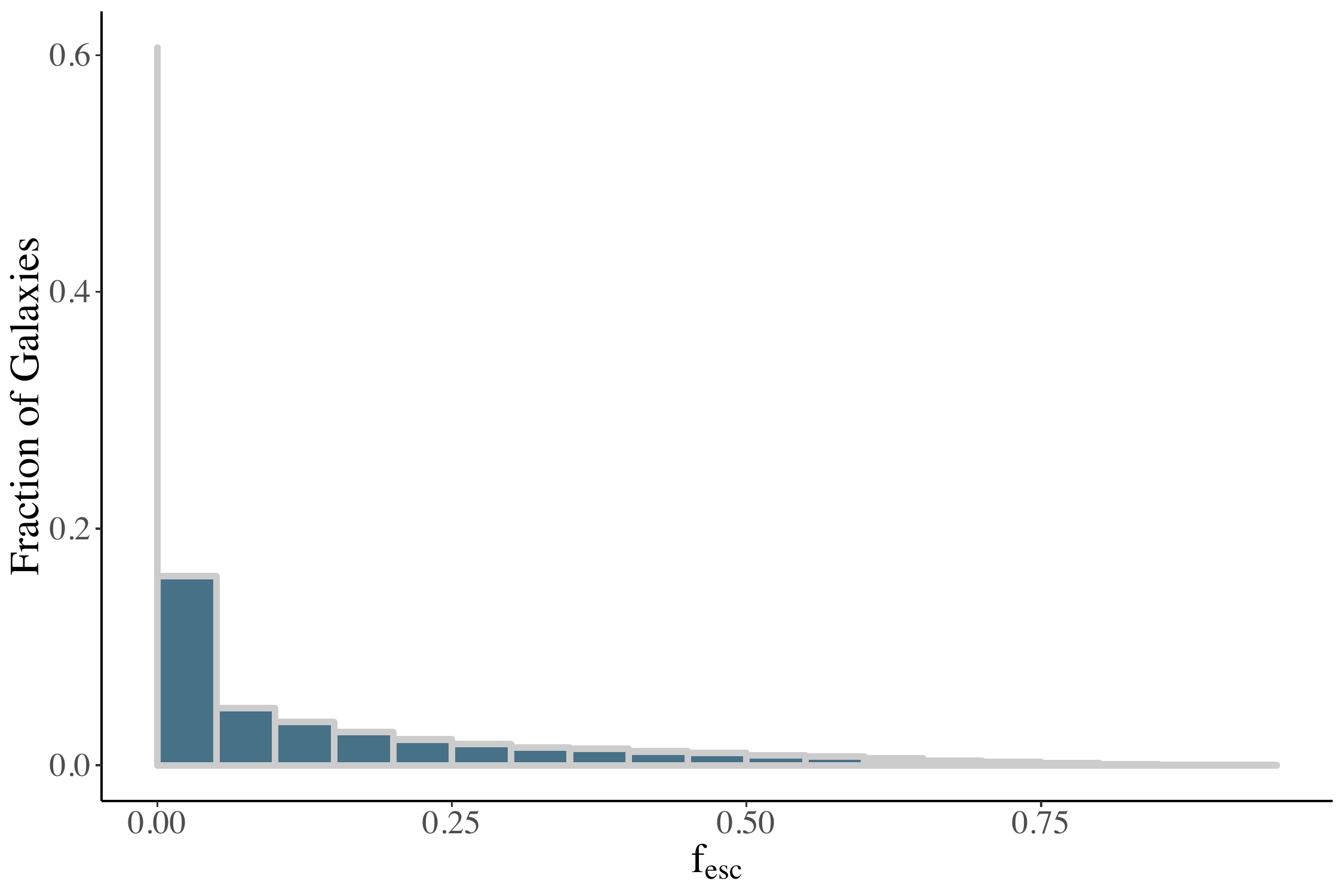}
\caption{Histogram of photon escape fraction. The leftmost narrow grey bar represents $f_{esc} = 0$ and encloses $\sim 60 \%$ of the sample. The  other 40 \% of galaxies with $f_{esc} > 0$ are displayed in bins of $\Delta f_{esc} = 0.05$. }
\label{fig:hist_fesc}
\end{figure}

The retrieved data consists of $75,683$ galaxies, with a  particle resolution of $M_{sph} = 1250 M_{\odot}$ and dark matter particle mass of $ M_{\rm{DM}}= 6246 M_{\odot}$,   in the redshift range $6 < \emph{z} < 25$, comprising halo masses in the range [$1.5\times 10^{6}- 5.8 \times 10^{9}$]~$M_\odot$. For each object, the corresponding $f_{esc}$ can be evaluated. The distribution of $f_{\mathrm{esc}}$ for these galaxies is displayed in Figure \ref{fig:hist_fesc}.  It is striking that no photons are able to escape in close to 60\% of the galaxies, which thus have $f_{\rm esc}=0$\footnote{All haloes with $f_{\rm esc} \leq 10^{-3}$.}, a behaviour intractable by Gaussian (a.k.a. normal)  approximations.

We study the behaviour of $f_{esc}$ through a set of seven variables representing different galaxy properties: {\it i}) the  stellar mass content ($M_{\star}~[M_{\odot}]$); {\it ii}) the halo virial mass ($M_{200}~[M_{\odot}]$); {\it iii}) the specific star formation rate (sSFR $[Gyr^{-1}]$), i.e. the star formation rate normalized by the stellar mass; {\it iv}) the baryonic fraction,  $\displaystyle f_b = \frac{M_{\star} + M_{gas}}{M_{200}}$, with $M_{\rm gas}$ being the gas mass inside the halo; {\it v})  the dark matter halo spin ($\lambda$); {\it vi}) the  total number of (HI) ionising photons that is being produced per second in the halo ($Q_{\sc HI}~[s^{-1}]$);
{\it vii}) the gas clumping factor, $C = \langle \rho^2 \rangle /\langle \rho \rangle^2$, where $\rho$ is the gas density averaged over all gas particles in the halo. 
The galaxies are analysed altogether stacking all snapshots below redshift 25.
Unless stated otherwise, throughout the analysis a logarithm transformation is applied to  all variables, except $f_b$, in order to deal with their wide range variability. Specifically, for sSFR  a log-modulus transformation $\displaystyle L({\rm sSFR}) = sign({\rm sSFR})\times \log(\|{\rm sSFR}\| + 1) \equiv \log({\rm sSFR}+ 1)$ is applied to account for the wide variance and  presence of zeros.  Hereafter, we use a clean notation for simplicity, i.e. without the use of the $\log$ terms and units.

\begin{figure*}
\centering
\includegraphics[width=\linewidth]{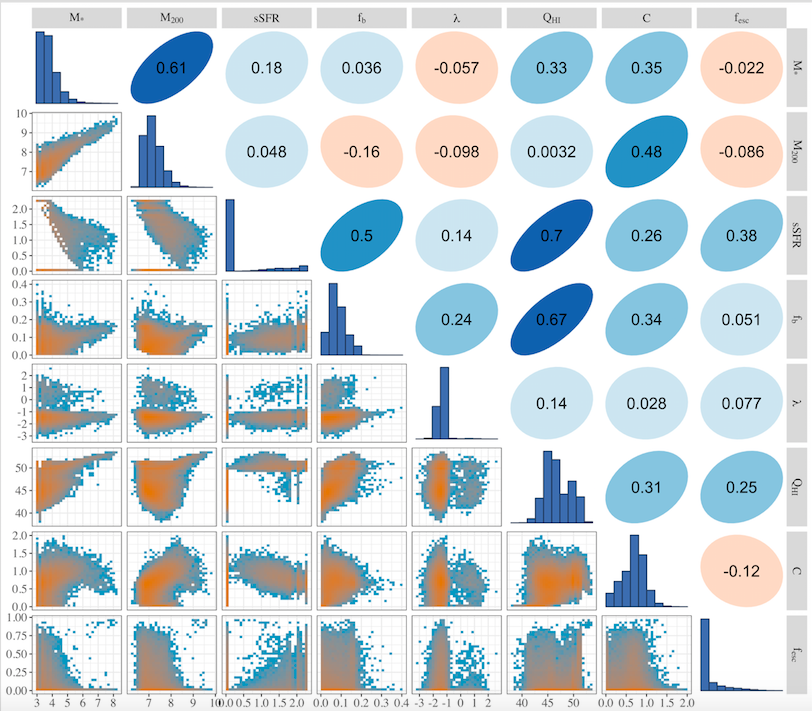}
\caption{Individual and pairwise representations of the galaxy properties. Histograms (units not shown) are given in the main diagonal. The lower diagonal represents a smooth scatter plot between each pair, whereas the upper diagonal gives the corresponding pairwise Spearman correlation coefficients.}
\label{fig:all}
\end{figure*}

Figure \ref{fig:all} provides a glimpse of the general data structure using univariate and pairwise diagrams of the transformed variables. The distribution of each galaxy property is given in the  main diagonal. The lower triangle  displays a smoothed scatter-plot for each pair of properties and the upper triangle gives the corresponding pairwise Spearman correlation coefficients. A visual inspection suggests that $f_{\mathrm{esc}}$ is most strongly related to sSFR, followed by $Q_{\sc HI}$. This can be explained noting that star formation must happen within a galaxy in order to produce ionizing radiation, and its escape into the IGM strongly depends on the total amount of ionising photons produced per second in the halo. 
However, with such complex and highly non-linear relationships among the various properties, a fully regression model is needed to quantify to what extent galaxy characteristics can serve as proxies for the level of $f_{\mathrm{esc}}$. 

Following the reasoning above, we can hypothesize that $f_{\mathrm{esc}}$ is determined by two processes, the first one defining a probabilistic threshold above which the photons are able to escape the halo, and a second one which determines the value  of $f_{\mathrm{esc}}$ once ionising photons have escaped. Hurdle models, also known as two-part  models, can take these two processes into account simultaneously, and thus they are natural candidates to use in our analysis.

\section{Statistical model formulation}
\label{sec:bbgamh}
\begin{figure}
\centering
\includegraphics[width=\linewidth]{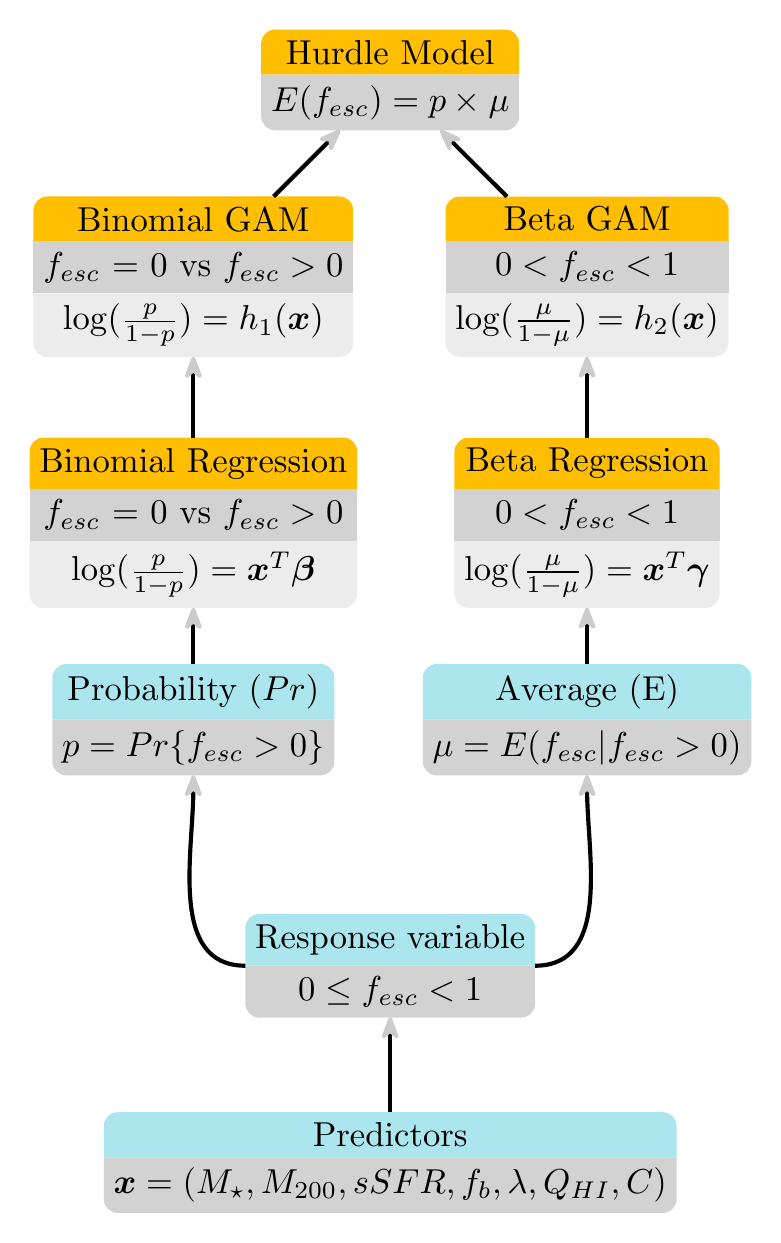}
\caption{A visual representation of the hurdle statistical model used to analyse the $f_{esc}$ dataset. From the bottom, the first two layers represent the galaxy properties that act as predictors and the response variable, $f_{esc}$.
There are two independent physical processes that control the data generating process of $f_{esc}$. The first process defines, $p$, the probability that $f_{esc} > 0$. Whereas the second process determines the fractional data, $f_{esc}$ when $f_{esc} > 0$. The first process, represented in left side of the diagram, is a $0$/$1$ process indicating whether $f_{esc} > 0$ and approximated by a binomial regression with logistic link. Whereas  logistic regression assumes linearity on the logit scale, a binomial generalized additive model (GAM)  drops this linearity assumption to allow for more complicated relationships between $f_{esc}$ and the predictors. Similarly, on the right side of the diagram, a beta GAM with logistic link is fitted to non-zero $f_{esc}$. Finally, a hurdle model sits on the top layer to combine information from both fitted models to wholly analyse $f_{esc}$. }
\label{fig:chart}
\end{figure}

Within this section we guide the reader through the construction of a customized statistical model one piece at a time. In order to motivate the reasoning behind the final model, we provide an overview of linear models (for Gaussian or normally distributed data), clearly stating their assumptions and limitations. We then discuss generalized linear models (GLMs) including  binomial regression (for binary data) and beta regression (for fractional data). Next, we introduce hurdle models that are capable of analysing data composed of a mixture of distributions. Finally, we present another layer of complexity suitable for describing non-linear relationships, the so-called  generalized additive models (GAMs).  While this will be done in the context of modelling $f_{esc}$, the underlying goal is to provide a general guideline to build customized statistical models for general astronomical purposes. In Figure \ref{fig:chart} a visualization of the statistical model used to analyse $f_{esc}$ is presented. Sections \ref{sec:LM}  to \ref{sec:GAM} provide the theoretical framework  to justify the model depicted in Figure \ref{fig:chart}.  Therefore, we refer directly to Section~\ref{sec:Final} the readers more  interested in the physical intuition behind the statistical model rather than in its mathematical foundations.

%
\subsection{Linear Models}
\label{sec:LM}

Albeit the assumptions of linear models can be restrictive, they have been successfully used in many astronomical applications \citep[e.g.][]{Isobe1990,Feigelson1992,Kelly2007,Sereno2016}. For this reason, and to understand better why they are not directly applicable to study the complex nature of the escape fraction, it is instructive to discuss them in further details.
We clearly specify the structure of linear models in both scalar and matrix forms and what is being estimated from the data. The exact form of these estimators is given along with their statistical properties.   

Linear models are powerful tools to study the relationship between a response variable and predictors (covariates). Consider the linear model, 
      \begin{align}
      &y = \beta_0 + \beta_1 x_1 + \beta_2 x_2 +\hdots +\beta_k x_k + e,\notag \\
      &E(e) =0, \quad var(e) = \sigma^2,
      \end{align}
where the response variable (or the outcome) $y$ represents an observable\footnote{Observable variable in statistical parlance means a variable for which we have a measurement information, which in astronomy can be either a real observation or simulated data.} random variable (e.g. $f_{esc}$), $\{1, x_1, x_2, \hdots , x_k\}$ constitutes a set of $p=k+1$ predictors ( e.g. Virial mass, baryonic fraction, ...) associated with the unknown regression parameters $\{\beta_0, \beta_1, \beta_2, \hdots , \beta_k\}$, $\sigma^2$ is an unknown positive parameter, $e$ is an unobserved random variable (error) with mean $0$ and variance $\sigma^2$, and $E(.)$ and $var(.)$ define the expected value 
and variance operators respectively. In this setting, the $x$'s are considered fixed known quantities after they have been sampled.   

Using basic properties of expectations and variances, the expected value (or mean) of $y$ is $\displaystyle E(y) = \beta_0 + \beta_1 x_1 + \beta_2 x_2 +\hdots +\beta_k x_k$, with $\displaystyle var(y) =\sigma^2$. The mean of $y$ is a weighted sum of the predictors. Theoretically speaking, the regression parameter $\beta_j$ represents the increase in the average of $y$ for each one unit increase in $x_j$ while holding the other predictors in the model constant. Notice that the variance is constant and does not depend on the mean.    

Lets formulate this model in matrix notation. Suppose that the data have been collected in pairs $(y_i,\vect x_i)$ for $i =1, \hdots , n$,  where $n$ is the sample size and $\displaystyle \vect x_i^T =  \begin{bmatrix} 1  & x_{i1} & x_{i2} \hdots & x_{ik}  \end{bmatrix}$. Then $\displaystyle y_i = \vect x_i^T \vect \beta + e_i$. It is typically assumed that $e_1$, $e_2$, $\hdots$, $e_n$ are uncorrelated random variables and thus $y$'s are also uncorrelated. 
This model can be represented in matrix notation,
      \begin{equation}
      \vect{Y} = \vect{X} \vect \beta + \vect e, \quad E(\vect e) =
      \vect{0}, \quad \quad Cov(\vect e) = \sigma^2 \vect I,
      \label{eq:modelone}
      \end{equation}
 where $\vect{Y}$ is an $n\times 1$ random vector of the response values $\displaystyle \begin{bmatrix}  y_{1} & y_{2} \hdots & y_{n} \end{bmatrix}^T$, $\vect{X}$ is an $n \times p$ known model matrix $\displaystyle \left[\vect x_i^T\right]$,
$\vect \beta$ is a $p \times 1$ vector of the unknown regression parameters $\displaystyle \begin{bmatrix}  \beta_{0} & \beta_{1} \hdots & \beta_{k} \end{bmatrix}^T$, $\vect I$ is an identity matrix of size $n$, $\vect{e}$ is an $n \times 1$ random vector of the unobservable random errors $\displaystyle \begin{bmatrix}  e_{1} & e_{2} \hdots & e_{n} \end{bmatrix}^T$ and $Cov(.)$ is the covariance operator. Hence, $E(\vect Y) = \vect X \vect \beta$ and $Cov(\vect Y) = \sigma^2 \vect I$. We assume throughout that $\vect X$ is a full rank matrix, i.e. rank of $\vect X$ equals $p$. This assumption is not necessary for linear models to work and it can be relaxed. The main goal is to gain knowledge about the predictors effects on the response variable through through estimation of $\vect \beta$ and hence $E(\vect Y)$,  based on the available data. 

A standard approach to estimate $\vect \beta$  is the ordinary least squares (OLS) method, which seeks to find $\vect \beta$ that minimizes $ (\vect Y -\vect X \vect \beta)^T (\vect Y - \vect X \vect \beta) = \vect e^T \vect e = \sum_{i=1}^{n} e_i^2$. The objective function is simply the sum of the squared errors. It can be proven that $ \hat {\vect \beta} =  \left(\vect X^T \vect X \right)^{-1} \vect X^T \vect Y$ is the unique solution for this optimization problem and denoted by the least squares estimate of $\vect \beta$. It can also be shown that  $E(\hat {\vect \beta} ) =\vect \beta $ and $Cov (\hat {\vect \beta} ) =\sigma^2 \left(\vect X^T \vect X \right)^{-1}$.

$E(\hat {\vect \beta} ) =\vect \beta $ means that $\hat{\vect \beta}$ is an unbiased estimator of $\vect \beta$, i.e. $\hat{\vect \beta}$ is correct on average. Furthermore, following the Gauss-Markov theorem \citep{christensen2011}, $\hat {\vect \beta}$ is the best linear unbiased estimator of $\vect \beta$, meaning that $\hat {\vect \beta}$ has the smallest variance among all linear functions of $\vect Y$ that are unbiased of $\vect \beta$. Moreover, if we further assume that $\vect e$ follows a multivariate normal distribution, then $\hat {\vect \beta} $ is the maximum likelihood estimator (MLE) and the best unbiased estimator. In this case, $\hat {\vect \beta} $ has a multivariate normal distribution as well. Identically, the previous results apply to any linear function of $\hat{\vect \beta}$. Along with the normality assumption equation \ref{eq:modelone} is denoted by Model \ref{eq:modelone}.         
      
  The predicted values (or the fitted values) are defined by $ \hat {\vect Y} = \vect X \hat{\vect \beta}$. An unbiased estimator of $\sigma^2$ is $\displaystyle \hat \sigma^2 = (\vect Y -\hat {\vect Y})^T (\vect Y - \hat {\vect Y} ) (n-p)^{-1}$. It can be proven that $(n-p)\hat \sigma^2$ has a $\chi^2$ distribution with $n-p$ degrees of freedom and independent of $\hat {\vect \beta}$. These properties along with $\hat{\vect \beta}$ properties described above allow to develop statistical tests and confidence regions around $\vect \beta$. 
  
 $\hat{\vect \beta}$ (or any estimator) is a random variable and if, hypothetically, another set of data has been obtained, the observed values of $\hat{\vect \beta}$ most likely will be different and in some cases can be very different. Therefore, it is essential to associate the estimates with a measure of uncertainty. We already have established that $\hat{\vect \beta}$ has a certain form of multivariate normal distribution. This fact can be exploited to construct region of plausible values of $\vect \beta$. For example, $\beta_1$ is estimated by $\hat \beta_1$ which is the second element of $\hat {\vect \beta}$. $\hat \beta_1$ has a normal distribution with mean $\beta_1$ and variance $\sigma^2 \times d_{22}$, where $d_{22}$ is the second element on the diagonal of $(\vect X^T \vect X)^{-1}$. By using normal distribution properties, rearranging some terms and replacing $\sigma^2$ by $\hat \sigma^2$, a $95\%$ confidence interval for $\beta_1$ is given by $\hat \beta_1 \pm 2 \times \hat \sigma \times \sqrt{d_{22}}$. More accurately, $2$ should be replaced by $t_{(0.975,n-p)}$, the $97.5\%$ percentile of a $t$ distribution with $n-p$ degrees of freedom. Statistical tests will be briefly discussed later. 
  
A main interest lies in estimating the average of $y$ at a specific combination of the predictors values stored at a $p\times 1$ vector, $\vect x_0$. This quantity is $E(y|\vect x_0) = \vect x_0^T \vect \beta$ and estimated by $\vect x_0^T \hat {\vect \beta} $. Since it is a linear function of $\hat {\vect \beta}$, $\vect x_0^T \hat {\vect \beta} $ is the best unbiased estimator of $\vect x_0^T \vect \beta$ and has a normal distribution with  $ var(\vect x_0^T \hat {\vect \beta}) = \vect x_0^T Cov (\hat {\vect \beta} ) \vect x_0 = \sigma^2 \vect x_0^T \left(\vect X^T \vect X \right)^{-1} \vect x_0$. Similar to above, 
\begin{equation}
\displaystyle \vect x_0^T \hat {\vect \beta} \pm 2 \times \hat \sigma \times  \sqrt {\vect x_0^T \left(\vect X^T \vect X \right)^{-1} \vect x_0}
\end{equation}
is a $95\%$ confidence interval for $\vect x_0^T \vect \beta $. 

 For space reasons, we could not discuss many important and critical aspects of linear models such as model diagnostics, weighted least squares, transformations and variable selection. We suggest the reader to consult \citet{christensen2011} for a comprehensive treatment on linear models, and \cite{kutner2005} for a detailed introduction.       

An important point to notice is that Equation \ref{eq:modelone} is considered linear because $E(\vect Y) = \vect X \vect \beta $  is a linear function of $\vect \beta$, but in principle, the model matrix $\vect X$ can contain quadratic, cubic or any polynomial function of the predictors. As from the scatter plots in figure \ref{fig:all} it is apparent that the relationship between $f_{esc}$ and halo properties cannot be described solely by linear terms, one could thus try  a more flexible model for $f_{esc}$ by adding quadratic terms in the form:   
  \begin{align}     
    f_{esc,i} &= \beta_0 +  \beta_1 M_{\star,i} + \beta_2 M_{\star,i}^2 + \beta_3 M_{200,i} +  \beta_4 M_{200,i}^2  + \notag \\
    &\hdots + \beta_{14} C^2_i  + e_i,
  \label{eq:modelthree}
  \end{align}   
where $i=1, \hdots , n$ refers to each of the $n=75683$ galaxies, $\left(e_i\right)_{i=1}^{n}$ has a multivariate normal distribution with mean $\vect 0$ and covariance $\sigma ^2 \vect I$, and $\vect X$ has $n$ rows and $p = 15$ columns (one for each  regression parameter)\footnote{Practically speaking, $\beta_1$ can no longer be interpreted as the increase in the average of $f_{esc}$ for each one unit increase in $M_{\star}$ while holding the other predictor variable in the model constant, since one cannot change $M_{\star}$ while keeping $M_{\star}^2$ fixed.}. Despite being relatively better suited for more complex relationships, this model is still restrictive and has the following serious drawbacks:

    \begin{enumerate}
    	\item While linear models are suited for Gaussian or normally distributed data, the distribution of $f_{esc}$ is very far from being normal (see figure \ref{fig:hist_fesc}).  A classical remedy for this problem is to utilize data transformation, i.e analysing a function of $y$ rather than $y$ itself. Due to the excess amount of zeros in the data, it is clear that any data transformation cannot achieve normality. For example, a log-transformation, after adding a positive constant equal to $0.001$, would just simply change the location of the long spike from $0$ to $-6.908$ and cannot cancel the discrete nature of the data.
        
\item While $f_{esc}$ can only take on values between $0$ and $1$, this model assumes that $f_{esc}$ is unbounded, and can thus produce predictions that have non-physical values, i.e. negative or greater than 1.  

 \item It assumes that the $f_{esc}$ variance is constant and independent of the mean, whereas figure \ref{fig:all} suggests that the variance is not fixed, but instead oscillates  across the predictors (i.e. the galaxy properties) space. 
 \item The relationship between $f_{esc}$ and the predictors is highly non-linear. It cannot be put in a pre-specified form and we cannot expect quadratic terms or even higher order terms to adequately model such complex relationships. Moreover, the degree of complexity is not the same across different predictors.  
    	\item As mentioned before, there may be  a mixture of two physical processes generated by two probabilistic models that control the behaviour of $f_{esc}$. Clearly the suggested model cannot adapt to this mixture, as an appropriate modelling must be able to take those two processes into account simultaneously. Besides $E(f_{esc})$, we also wish to model the probability of a given galaxy to trigger the escape of photons.
 \end{enumerate}
 
Next we seek to address these five limitations, ill-suited for chaotic systems,  to be able to model more appropriately the complexity of $f_{esc}$.

\subsection{Generalized Linear Models}
\label{sec:GLM}

The goal of this section is to lay the basis for addressing the limitations of the linear models discussed previously, and to guide the reader through the construction of the more elaborate model presented in section \ref{sec:Final}.
To this aim, we give a short but solid introduction to generalized linear models (GLMs),
touching upon the most fundamental aspects of GLMs, including their structure and assumptions, estimation method, statistical properties of the estimators, confidence intervals for quantities of interest, statistical tests and model selection criteria. 

By extending some of the linear model assumptions, GLMs \citep[introduced by][]{nel72} can apply to data sets for which linear models are ill suited. In fact, GLMs contain Model \ref{eq:modelone}  as a special case. GLMs are pretty rich and can accommodate various types of data sets for which the response variable can be binary, count, symmetric, highly skewed, ordinal or nominal. Basically, assumptions of GLMs can be different from those in Model \ref{eq:modelone} in three aspects: first, the distribution of the response variable is not necessarily normal; second, the variance of the response variable can be a function of the mean; and finally, a non-linear relationship between $E(y)$ and $\vect x^T \vect \beta$ is possible. Both GLMs and Model \ref{eq:modelone} assume that the responses are independent random variables.  

GLMs are centred around the concept of the so called link function and the distribution of the response variable being a member of the exponential family. The link function expresses the relationships between $E(y)$ and the linear predictor $\vect x^T \vect \beta$. The distribution which depends on a parameter vector $\vect \theta $ belongs to the exponential family if its probability function can be expressed in a specific form
\begin{equation}
f(y;\vect \theta) = h(y)A(\vect \theta)\exp\{W(\vect \theta)^TT(y)\}, 
\end{equation}
 where $h$, A, W and T are known functions.  
Normal, binomial, Poisson and Gamma distributions are examples of exponential family distributions. Formally, GLMs assume that: 
\begin{itemize}
      \item $\left(y_i\right)_{i=1}^{n} $ forms an independent set of random variables each of which has the same distribution that is a member of the exponential family;
      \item $g(E(y)) = \vect x^T \vect \beta$, where $g(.)$ is a monotonic and differentiable function and denoted by the link function. Notice that $E(y) = g^{-1}(\vect x^T \vect \beta)$. For a non-negative random variable $y$,  a square root link function, for example, assumes that $\sqrt{E(y)} = \vect x^T \beta$ or equivalently  $E(y) = (\vect x^T \beta)^2$.
\end{itemize}  
      The linear structure in the link function allows to extend to GLMs statistical methods developed for linear models  \citep{christensen2011}. The distribution of $y$ determines the relationship between the mean and the variance. Model \ref{eq:modelone}  described in the previous section assumes an identity link, i.e. $g(\mu) =\mu = \vect x^T \vect \beta$, and the mean is independent of the variance.  A binomial regression with logistic link (denoted by logistic regression and discussed in the next section)  and a Poisson regression with log link, i.e. $g(\mu) =\log \mu = \vect x^T \vect \beta$, are typically used when modelling binary responses and count data, respectively. For Poisson regression, the mean and the variance are identical. Gamma regression is suitable for skewed responses with constant coefficient of variation. The interpretation of the regression coefficients depends on the link function being used. 

In GLMs, the parameter vector $\vect \theta$ consists of the regression coefficient vector $\vect \beta$ and a dispersion parameter $\phi$ that is related to the response variance. For linear regression, $\phi = \sigma^2$, and for Poisson and binomial regressions $\phi = 1$. 

$\vect \theta = (\vect \beta, \phi)$ is usually estimated through the maximum likelihood method, which aims to find $\vect \theta$ that maximizes the log-likelihood function. The likelihood function for  $\vect \theta = (\vect\beta, \phi)$ is given by 
  \begin{equation}    
 L(\vect\beta, \phi)  \equiv f(y_1,\hdots y_n;\vect\beta, \phi ) = \prod_{i=1}^{n} f_i(y_i;\vect\beta, \phi ),
\end{equation}
where $f_i$ is the probability function of $y_i$. Apart from linear regression and some trivial cases, explicit optimization of the log likelihood, $l(\vect\beta, \phi) = \sum_{i=1}^{n} \log f_i(y_i;\vect\beta, \phi )$ is not possible. In contrast to Model  \ref{eq:modelone}, the MLE usually does not have an exact form. 
However, accurate approximation of the MLE, $\hat{\vect \theta} = (\hat{\vect \beta},\hat {\phi} )$, can be found using numerical methods such as iterative weighted least-squares routine via Newton-Raphson algorithm or Fisher's scoring method. The details of these algorithms can be found in \citet{Hil12}.
                  
     Asymptotically, $(\hat{\vect \beta},\hat {\phi} )$ has a multivariate normal distribution with mean equals to  $({\vect \beta}, {\phi} )$  and covariance matrix of
\begin{equation}
 \vect H(\vect \beta,\phi) = [-E(\vect G(\vect \beta,\phi))]^{-1},
 \label{eq:fisher}
\end{equation}
where $\vect G(\vect \beta, \phi)$ is a $p+1$ square matrix containing the second order partial derivatives of the log-likelihood with respect to $\vect \beta$ and $\phi$. $\vect H$ is commonly estimated by $ \hat {\vect H} = [-\vect{G}(\hat{\vect \beta},\hat{\phi})]^{-1}$. Practically speaking, for large $n$, $\hat{\vect \beta}$ has a small bias and approximately normally distributed. The standard errors of $\hat \beta_j$'s are estimated by the first $p$ diagonal elements of $\hat {\vect H}$. Using multivariate normal properties, asymptotic tests and interval estimations of any linear function of $\vect \beta$ can immediately follow. 

The average of $y$ at $\vect x_0$ is $E(y|\vect x_0) = g^{-1}(\vect x_0^T \vect \beta)$ and estimated by $g^{-1}(\vect x_0^T \hat{\vect \beta})$. An asymptotic $95\%$ confidence interval of $\vect x_0^T \vect \beta$ is given by $\displaystyle \vect x_0^T \hat {\vect{\beta}} \pm 2 \ \sqrt{\vect x_0^T \hat {\vect \Tau} \vect x_0} $, where $\hat {\vect \Tau}$ is the matrix $\hat{\vect H}$ excluding the last column and the last row which correspond to $\hat \phi$. Let $(a,b)$ be an asymptotic $95\%$ confidence interval for $ \vect x_0^T \vect \beta$. Then, depending on the link function being used, $\left(g^{-1}(a),g^{-1}(b)\right)$ or $\left(g^{-1}(b),g^{-1}(a)\right)$ is an approximate $95\%$ confidence interval for $E(y|\vect x_0) = g^{-1}(\vect x_0^T \vect \beta)$. 

Wald tests \citep{Wald43,silvey1959}, which depend on the sampling distribution of $\hat{\vect \beta}$, are frequently used to test the hypothesis that $\beta_j$ is $0$ or to simultaneously test the hypothesis that multiple components of $\vect \beta$ are zeros. For example, assume that we are interested in testing the hypothesis that $\beta_1 = \beta_2 = 0$, i.e the first two predictors have a null effect on the average of $y$ after adjusting for the other predictors, and let $\hat{\vect V}$ be the estimated covariance matrix of $[\hat \beta_1  \ \hat \beta_2]^T$ extracted from the second row and second column, and the third row and the third column of the matrix $\hat {\vect \Tau}$. The Wald test statistic is $\displaystyle W = [\hat \beta_1  \ \hat \beta_2] \hat {\vect V}^{-1} [\hat \beta_1  \ \hat \beta_2]^T$. $W$ has a $\chi^2$-distribution with $2$ degrees of freedom ($\chi^2_2$) if the assumption $\beta_1 = \beta_2 = 0$ is correct. If $r$ components of $\vect \beta$ are involved in this hypothesis, then the degrees of freedom are $r$. If the observed value of $W$ is larger than what we would expect from a $\chi^2_2$-distribution then perhaps the null hypothesis $\beta_1 = \beta_2 = 0$ is not correct. Formally speaking, the null hypothesis is rejected at a level of significance $0.05$, if the observed value of $W$ is greater than the $95\%$ percentile of a $\chi^2_2$-distribution. $F$ distribution can be used instead if the distribution of $y$ contains $\phi$,  the dispersion parameter.     

Alternatively, the likelihood ratio test (LRT) can be conducted as follows. Let $\hat{\vect \theta}_0$, and $\hat{\vect \theta}$ be the MLE for the parameters of model $M_0$ and model $M$ respectively where model $M_0$ is nested within (i.e. a special case) of model $M$. For the above example, $M_0$ will fit the same model as $M$ without the first two predictors. The hypothesis that asserts that model $M_0$ is true and $M$ is not needed, is rejected at $0.05$ significance level if the observed value of the test statistic $\displaystyle \lambda = -2 [l_{0}(\hat {\vect \theta}_0) - l(\hat {\vect \theta)} ] $ is greater than the $95\%$ percentile of a $\chi^2_r$-distribution, where $l_0(.)$  is the log-likelihood of $M_0$ and $r$ represents the difference in the number of parameters between the two models. This is a rich and widely used test and can be applied in many situations and not restricted to performing tests on $\vect \beta$. For the above example, Wald test and LRT are equivalent for normal linear regression and both yield similar results for other members of GLMs.   

In many cases there is a set of competing models which are not nested within each other and thus LRT cannot be used, e.g. two Poisson models with  different link functions. In this case, to choose between them, one can revert to model comparison criteria, such as the Akaike Information Criteria \citep[AIC;][]{Akaike1974}, or  
the Bayesian Information Criterion \citep[BIC;][]{BIC}, which are calculated as follows: 
\begin{eqnarray}  
        AIC &=&   -2 \ l(\hat {\vect \theta}) +2 \ (p+1), \nonumber\\
        BIC &=& -2 \ l(\hat {\vect \theta}) + \log n \ (p+1). 
        \label{eq:IC}
        \end{eqnarray}

When  AIC/BIC are used for model selection, the preference is given to the model with the lowest AIC/BIC. Although an increase in the log-likelihood is desired, 
it can artificially increase as an effect of the use of more parameters, then leading to  over-fitting. To mitigate such effect, the second term of the AIC and the BIC penalizes for the number of parameters in the model (the model complexity). We refer the reader to  \citet{mcc89} for an extended discussion on GLMs and their applications. See also \citet{dobson2010}  for an introductory approach.  

%

\subsubsection{Logistic Regression}  

This section introduces the formalism behind a particular case of GLMs, namely, logistic regression, which we will employ to model the probability that $f_{esc} > 0$ for a galaxy with a given set of properties. 

Logistic regression \citep[see e.g.][for a review]{hilbe2016practical,Hilbe2017} is very widely used by practitioners from different fields, such as ecology \citep{Pearce2000225} or medical sciences \citep{HeinzeSIM1047},  and has been increasingly  used in  astronomy, for example, to probe the likelihood of star forming activity in primordial galaxies \citep{deSouza2015AeC}, or to model the environmental effects in the presence/absence of active supermassive black holes \citep{deSouza2016MNRAS}. It generally aims to model binomial (binary) data.  

A binomial distribution describes a sequence of independent experiments (trials) each of which has only two possible outcomes $\{0,1\} $. 
If  $y\sim \textrm{Binomial}(N,p) $, then 
\begin{equation}
f(y;p) = \binom{N}{y} p^y (1-p)^{N-y} \quad {\rm for}~  y = 0,1,\hdots, N, 
\end{equation}
where $N$ represents the number of independent trials, and $p$ is the probability of success. $y$ basically counts the number of successes out of $N$ trials. $E(y) = N p$ and $var(y) = N p (1-p)$. For $N=1$, $y$ can only take two values, $0$ or $1$, and $E(y)$ is just the probability of success, $p$. 

In the specific case of interest here, we can think of the escape fraction as binary data which is either $f_{esc}=0$ or $f_{esc}>0$. We thus aim to estimate the probability of having a non zero escape fraction, i.e. $p = Pr\{f_{esc} > 0\}$.

Notice that assuming an identity link, $E(y) = \vect x^T \vect \beta$, can produce probabilities greater than $1$ or less than $0$. 
Logistic regression, instead, assumes the widely used logit (or $\log \text{odds}$) link, 
\begin{align}
&\log \frac{p}{1-p} = \vect x^T \vect \beta \quad \rm{or} \nonumber \\
&p = \left(1+\exp(-\vect x^T \vect \beta) \right)^{-1},
\end{align}
which forces $p$ to be between 0 and 1, as it should be.
The exponentiated regression coefficient,  $\exp(\beta_j)$,  represents the increase in the odds of success for each one unit increase in $x_j$,  while holding the other predictors constant. The log likelihood function for this model is given by,
\begin{equation}
l(\vect \beta)  \propto \sum_{i=1}^{n} \left\{y_i \log p_i  + (N_i-y_i) \log (1-p_i) \right\}.
\end{equation}
Numerical methods such as iterated weighted least squares are needed to produce $\hat {\vect \beta}$. For large $n$, the covariance matrix of $\hat {\vect \beta}$ is approximately $ (\vect X^T\vect D\vect X )^{-1}$ where $\vect D =  \text{diag}(N_i p_i(1-p_i))$. 

The estimated probability of success at $\vect x_0$ is then
\begin{equation}
 \hat p_0 = \left(1+\exp(-\vect x_0^T \hat {\vect \beta}) \right)^{-1}. 
\end{equation}
The large sample covariance matrix of $\hat {\vect \beta}$ is estimated by replacing $\vect D$ by $\hat{\vect D} = \text{diag}(N_i\hat p_i(1-\hat p_i))$. An asymptotic $95\%$ confidence interval for $p_0$ is given by $\left(1+\exp(-L) \right)^{-1} $ and $\left(1+\exp(-U) \right)^{-1} $ where $(L,U)$ is   
\begin{equation}
 \vect x_0^T \hat \beta \pm 1.96 \ \sqrt{\vect x_0^T (\vect X^T \hat{\vect D}\vect X )^{-1}\vect x_0}, 
\end{equation}
representing a $95\%$ confidence interval for $\vect x_0^T \vect \beta$ computed using normal approximation of $ x_0^T \hat {\vect \beta}$ distribution\footnote{
A less popular model for binomial data is the probit regression, in which the link function is given  by the inverse of the cumulative distribution function of standard normal distribution.}.
%
\subsubsection{Beta Regression}

This section introduces a model suitable to treat fractional data, which will be of particular interest to study the fractional values of $f_{esc}$. While logistic regression models the probability that $f_{esc} >0$ for a given galaxy, i.e. $p = Pr\{f_{esc} > 0 \}$, beta regression will be used to model the mean value of $f_{esc}$ for galaxies with $f_{esc} >0$, i.e. $E(f_{esc}|f_{esc} >0)$. 

 Beta distribution describes continuous random variables that fall naturally between $0$ and $1$, such as proportions, rates and concentrations. The beta probability function is characterized by two positive shape parameters, $a$ and $b$. The standard beta probability density function is given by,
\begin{equation}
f(y) = \frac{\Gamma(a+b)}{\Gamma(a)\Gamma(b)}y^{a-1}(1-y)^{b-1},\qquad 0 < y < 1.
\end{equation}
The mean, and variance of $y$ are,

\begin{align}
&E(y) = \frac{a}{a+b} = \mu,\notag\\ 
&var(y) = \frac{ab}{(a+b)^2(a+b+1)}.
\end{align}

\begin{figure}
\centering
\includegraphics[width=\linewidth]{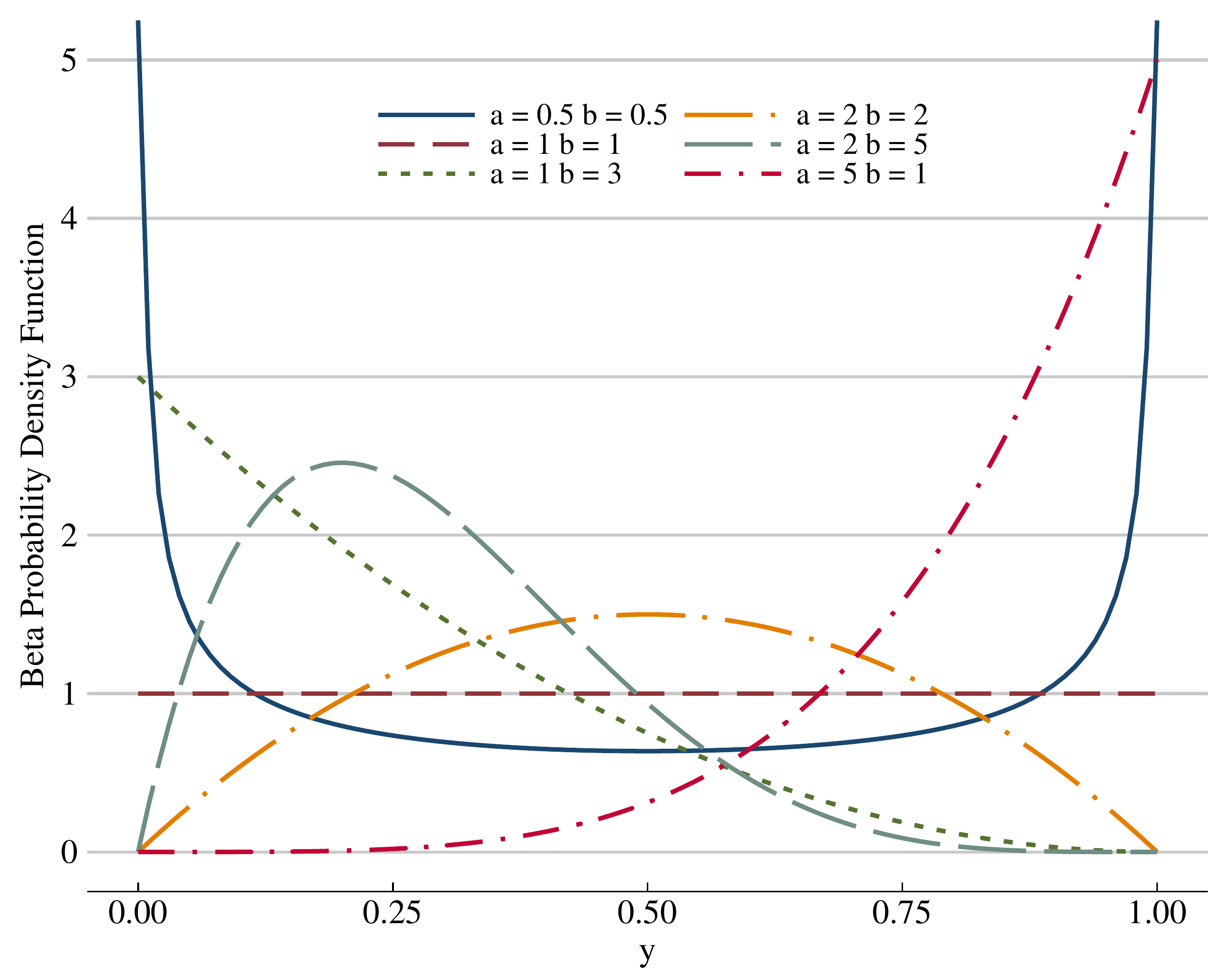}
\caption{Beta densities with different values of the shape parameters $a$ and $b$.}
\label{fig:beta_dist}
\end{figure}

 The variance is related to the mean through the relationship, $\displaystyle var(y)  = \frac {\mu (1-\mu)}{1+a+b}$ and consequently variance heterogeneity is implied. Figure \ref{fig:beta_dist} displays beta densities for multiple values of $a$ and $b$. It can be seen that depending on $a$ and $b$, beta distribution can take various shapes such as bell-shaped (normal like), U-shaped (bimodal), right skewed, left skewed and flat (corresponding to a uniform $(0,1)$ distribution when $a=b=1$). 

\citet{Ferrari2004} proposed beta regression to model a relationship between a unit interval $y$ and covariates, $\vect x$. First they suggested a different parametrization of beta distribution in terms of $\mu$ and $\phi = a +b $. The dispersion parameter $\phi$ has inverse relationship with $var(y)$. For this new parametrization
\begin{align}
&f(y;\mu,\phi) = \frac{\Gamma(\phi)}{\Gamma(\mu \phi)\Gamma((1-\mu) \phi)}y^{\mu \phi-1}(1-y)^{(1-\mu)\phi-1},\nonumber\\
& 0 < y < 1.
\label{pdf:beta}
\end{align}
Logit, probit and complementary log-log can be used as link functions. For logit link, $\displaystyle \log\frac{\mu}{1-\mu} = \vect x^T \vect \beta$. 
The log likelihood function can be expressed as,
\begin{dmath}
l(\vect \beta, \phi) = n \log \Gamma(\phi) - \sum_{i=1}^{n} \Gamma(\mu_i \phi) - \sum_{i=1}^{n} \Gamma((1-\mu_i) \phi) + \sum_{i=1}^{n} (\mu_i \phi -1) \log y_i + \sum_{i=1}^{n} ((1-\mu_i) \phi -1) \log (1-y_i).
\end{dmath}
Numerical methods can optimize  $l(\vect \beta, \phi)$. The average value of $y$ at $\vect x_0$, $\mu_0$, is estimated by 
\begin{equation}
\hat \mu_0 =\left(1+\exp(-\vect x_0^T \hat {\vect \beta}) \right)^{-1}.
\end{equation}
As described before for GLMs, finding a $95\%$ confidence interval for $\mu_0$ is made possible through maximum likelihood theory. 

\begin{figure*}
\centering
\includegraphics[width=0.45\linewidth]{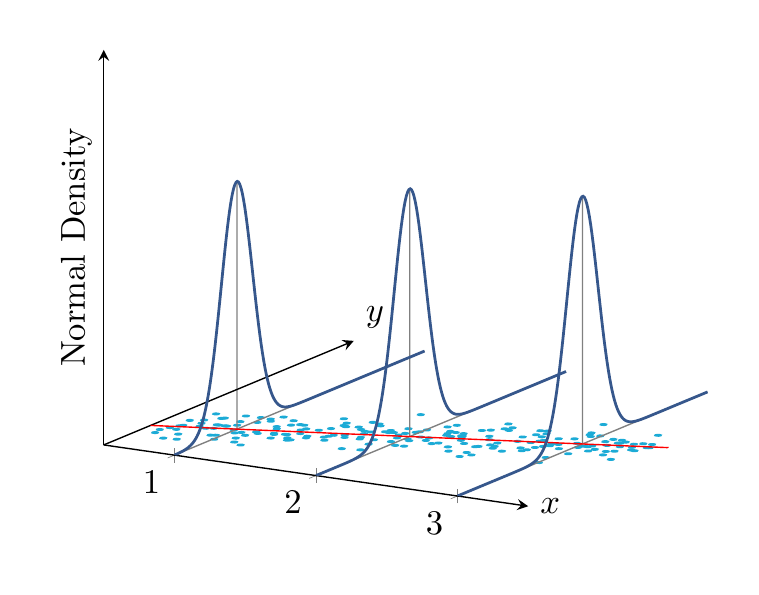}
\includegraphics[width=0.45\linewidth]{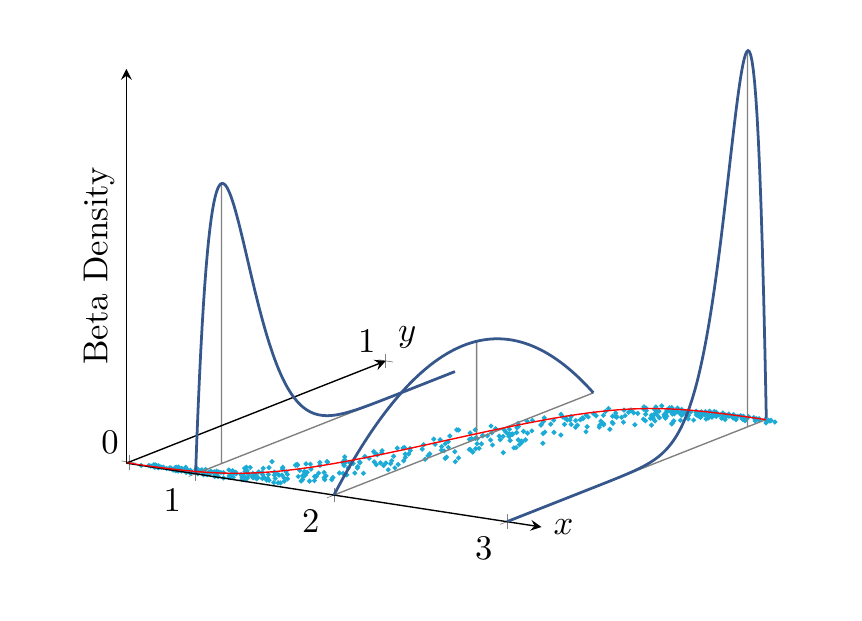}
\caption{ Two illustrative simulation studies where the left and right panels represent a linear regression and a beta regression with logistic link, respectively. The true mean regression curves are presented in red. The light-black vertical lines point towards the response mean.}
\label{fig:beta_reg}
\end{figure*}

Figure \ref{fig:beta_reg} displays two illustrative simulation studies where the left and right panels correspond to a  linear regression with normal errors and a beta regression, respectively. The true response mean regression functions are represented by the red curves. In the linear regression case, the variability of the response $y$ does not change with the mean and in fact stays constant. The normal density curves are exact replica of each other except for the location of the mean, which depends on the predictor $x$ through some linear relationship. In the beta case, both the variability of the response and the shape of the distribution depend on the mean. As the response mean increases, the shape of the distribution switches from right skewed to symmetric to left skewed. A logistic link is assumed for the beta regression justifying the non-linear relationship between $x$ and the response mean.        

If $y$ does not fall into a unit interval $(0,1)$ but rather into an interval $(c,d)$, $y$ can be easily transformed by applying simple transformation  $\displaystyle \frac{y-c}{d-c}$ \citep{Ferrari2004}.

 \subsection{Hurdle Models}
\label{sec:hurdle}

As noticed in Figure \ref{fig:hist_fesc}, about $60\%$ of galaxies have $f_{esc}=0$, indicating that photons on these galaxies were unable to reach the intergalactic medium. Once this happens, then $f_{esc}$ fractional and continuous values are obtained. As this unusual data distribution cannot be accommodated by GLMs, here we present hurdle models, that can take into account simultaneously the two processes determining $f_{esc}$, i.e. the first one defining a probabilistic threshold above which the photons are able to escape the halo, and the second one which determines the value of $f_{esc}$ once they have escaped.

To the best of our knowledge,  hurdle models (see \citealt{Hilbe2017} for a discussion on few simple examples) have never been applied to a real astrophysical problem. In this section, we present the technical and quantitative details that are necessary to fully understand such models for both discrete and continuous data, as is the case of $f_{esc}$, and we illustrate how they can be applied not only to $f_{esc}$, but also to other astronomical problems.

Zeros appear naturally in discrete distributions, such as binomial and Poisson distributions. Excess zeros occur when the number of observed zeros is more than expected under the sampling model. This phenomenon is not unusual in Astronomy. For example, one may be interested in estimating the number of Earth-like planets in  habitable zones in terms of the stellar system properties. Such a survey may lead to several systems without any candidate, suggesting that a mixture of two processes has generated the data. For instance one process to trigger the formation of Earth-like planets and another independent process that puts them or not in the habitable zone. As for the case of the escape fraction, GLMs are not appropriate to describe this phenomenon while hurdle models are.

A hurdle model (originally proposed for Poisson count data by \citealt{Mulla86}) is a two-component mixture model characterizing separately the sampling process of $\{y >0 \text{ vs. } y=0 \}$, denoted by $1_{y > 0}$, and the sampling process of $\{y$ given that $y>0 \}$, denoted by $y|y>0$. 
It thus consists of two stages: the first stage seeks a binomial model addressing whether $y$ is zero or positive; if $y>0$, the second stage is initialized to address the conditional distribution of $y$.  
Suppose that $w$ a non-negative discrete random variable that has a probability function $f(.)$ such that $f(0) < 1$ with mean $\mu$ and variance $\sigma^2$. The model assumes   
\begin{equation}
         Pr\{ y=j \} = 
         \begin{dcases} 
        1-p & ; j=0 \\
         \frac{p}{1-f(0)} f(j) & ; j > 0.  
        \end{dcases}
 \end{equation}
 It is straightforward to show that
 
 \begin{eqnarray}  
        E(y) &=&   \frac{p} {1-f(0)} \mu, \nonumber\\
        var(y) &=& \frac{p} {1-f(0)} \left[\sigma^2 + \frac{1-p-f(0)} {1-f(0)} \mu^2 \right].
        \label{eq:Hurdle}
        \end{eqnarray}
 For Poisson distribution, the mean and the variance are identical (i.e $\sigma^2 = \mu$). This can be seen as a restrictive assumption. Hurdle models can relax this assumption to allow for different relationships between the mean and the variance. If $w$ follows Poisson distribution, it can be shown that $\displaystyle var(y) = E(y) + C$. If $1- f(0) > p$, then $ C > 0 $ which allows for over-dispersion, i.e. $var(y) > E(y)$. Under-dispersion occurs, i.e. $var(y) < E(y)$, when $C  < 0 $.  $C=0$ if and only if $p=1-f(0)$ and thus $y$ has a Poisson distribution as $w$. Consequently, Poisson distribution can be seen as a special case of a hurdle model. Formal statistical tests, such as a likelihood ratio test, can be used to assess whether the Poisson model is adequate.

If $w$ is viewed as a  continuous random variable as the case of $f_{esc}$, then hurdle model assumes that $\displaystyle Pr\{0 \leq y < a \} = (1-p) + p \int_{0}^{a} f(w) dw $. Clearly, $1-p=Pr\{y=0\}$. The mean and variance are         
        \begin{eqnarray}  
        E(y) &=& p \mu, \nonumber\\
        var(y) &=& p\left[\sigma^2 + (1-p) \mu^2 \right].     
        \label{eq:meanvarhur}
        \end{eqnarray}
            
If $p=0$, $y$ is degenerate at $0$ and $p=1$ implies that the density of $y$ is simply $f$. The mean is smaller than $\mu$ since the zeros pull the mean down. A reduction in the variance is achieved, i.e. $var(y) < \sigma^2$, if $\displaystyle  \mu  < \frac{\sigma}{\sqrt{p}}$. 

As in GLMs, hurdle regression requires for both processes, $1_{y > 0}$ and $y|y>0$, the determination of response distributions, predictors and link functions. The predictors for both processes do not need to be the same. In other words, $\displaystyle g_1(p) = \vect x_1^T \vect \beta_1$ and $\displaystyle g_2(\mu) = \vect x_2^T \vect \beta_2$. It can be shown that the log-likelihood of $\vect Y$,   \begin{dmath}
 	l(\vect \beta_1,\vect \beta_2, \phi) = l_1(\vect \beta_1) + l_2(\vect \beta_2,\phi),
    \label{eq:hurdlelog}
 \end{dmath}
a sum of two log-likelihoods. This is an attractive property of hurdle models allowing to find the MLE of $(\vect \beta_1,\vect \beta_2,\phi)$ by optimizing $l_1$ and $l_2$ separately to yield $(\hat {\vect \beta}_1,\hat{\vect \beta}_2,\hat {\phi})$.  This decomposition would not be possible if $\vect \beta_1$ is assumed to have some relationship with $\vect \beta_2$. Such a decomposition allows to further complicate the structure of the model without adding computational difficulties. For instance, if the correlation between any two non-zero responses decreases as they fall apart in the space, then one can add spatial components to the model. Using exiting software packages, this hurdle model can be fitted by fitting a Binomial-GLM to  $1_{y > 0}$ and a spatial-GLM to $y|y>0$ separately given that the parameters for both models are not the same.        

The terms $\hat{\vect \beta}_1$ and $(\hat{\vect \beta}_2,\hat {\phi})$ are statically independent as a result of $\displaystyle \frac{\partial l^2}{\partial \vect \beta_1 \partial \vect \beta_2} = 0$. The large sample distribution of $(\hat {\vect \beta}_1,\hat{\vect \beta}_2,\hat {\phi})$ is a multivariate normal with mean $(\vect \beta_1,\vect \beta_2,\phi)$, and   covariance matrix consisting of four blocks. The two blocks over the diagonal are computed as in equation \ref{eq:fisher} and the off-diagonal blocks are zero matrices. Statistical inferences regarding $p$ and $\mu$ can be conducted separately and as usual. Also, it is easy to build a Wald test or a LRT to simultaneously test if a set of predictors can be removed from both components altogether. Let $T_1$ and $T_2$ be two statistics (Wald tests or LRTs as defined in Sec. \ref{sec:GLM}) to test  $\beta_{11} = \beta_{12} = 0$ and $\beta_{21} = \beta_{22} = 0$, respectively. Then $T_1 + T_2$ has a $\chi^2_{4}$-distribution when $\beta_{11} = \beta_{12} = \beta_{21} = \beta_{22} = 0$, which implies that $x_1$ and $x_2$ have null effects on $p$ and $\mu$ after adjusting for other predictors. This hypothesis is rejected at the significance level $0.05$ if the observed value of $T_1 + T_2$ exceeds the $95\%$ percentile of $\chi^2_{4}$.   

According to equation (\ref{eq:meanvarhur}), the average value of $y$ at $\vect x_0$ is $E(y|x_0) = p_0\mu_0$. An estimate of this quantity is 
\begin{equation}
\hat m_0 = \hat p_0 \hat \mu_0  = g_1^{-1}(\vect x_{01}^T \hat {\vect \beta}_1) \times g_2^{-1}(\vect x_{02}^T \hat {\vect \beta}_2),
 \label{eq:Hurdle}
\end{equation}
where $\vect x_{01}$ and $\vect x_{02}$ are extracted from $\vect x_0$. Since $\hat {\vect \beta}_1$ and $\hat{\vect \beta}_2$ are MLEs and statistically independent, $E(\hat m_0)$ is approximately $E(y|x_0)$. Furthermore, for large $n$, $\hat {\vect \beta}_1$ and $\hat{\vect \beta}_2$ have a multivariate normal distribution, and one can use parametric bootstrapping to compute approximate confidence intervals of $E(y|x_0)$. Alternatively, delta method can be used to calculate the estimated standard error of $\hat m_0$ and thus approximate confidence intervals can be computed as follows. 

The estimated standard errors (SE) of $\hat p_0$ and $\hat \mu_0$ can be approximated based on Taylor series expansion. When assuming logit link for the binomial part of the process $1_{y > 0}$, we obtain
\begin{equation}
\text{SE}(\hat p_0) \approx \frac{\exp(-\vect x_{01}^T \hat {\vect \beta}_1)}{\left(1+\exp(-\vect x_{01}^T \hat {\vect \beta}_1) \right)^{2}} \sqrt{\vect x_0^T (\vect X^T \hat{\vect D}\vect X )^{-1}\vect x_0}. 
\end{equation}
The estimated standard error of $\hat \mu_0$, $\text{SE}(\hat \mu_0)$, can be computed using similar techniques. Then using Delta method,
\begin{equation}
\text{SE}(\hat m_0) \approx \sqrt{\text{SE}(\hat p_0) ^2 \hat \mu_0^2  + \text{SE}(\hat \mu_0)^2 \hat p_0^2 + \text{SE}(\hat p_0)^2 \text{SE}(\hat \mu_0)^2},
\label{eq:SE}
\end{equation}
and $\displaystyle \hat m_0 \pm 2\ \text{SE}(\hat m_0) $ is an approximate $95\%$ confidence interval of $E(y|x_0)$. The accuracy of this interval depends on $n$ being large. 

In summary, in this section we presented hurdle models that will be used to model $f_{esc}$ through a combination of two processes. In the next section, we will relax the linearity assumption to account for the complex relationship between $f_{esc}$ and galaxy properties (see figure \ref{fig:all}).

%
 \subsection{Generalized Additive Models}
 \label{sec:GAM}
As mentioned above, in this section we seek to address the linearity assumption that deems inappropriate according to figure \ref{fig:all}.
So far we have been assuming that 
\begin{equation}
g(E(y)) = \vect x^T\vect \beta \label{eq:linear}. 
\end{equation}
This functional relationship takes a specific linear formula and is completely specified except for the unknown regression vector $\vect \beta$. Generally speaking, $g(E(y))$ is unknown, i.e. $g(E(y)) = h(\vect x)$ for some unknown regression function $h$. GLMs assume that $h(\vect x)$ can be approximated by $ \vect x^T \vect \beta$. This linear approximation, despite being simple, has been proven valuable for many applications. Also, as we have demonstrated, estimation and inferences are straightforward. 
     
     For many instances, the functional formula in equation \ref{eq:linear} can be quite restrictive. It may not account properly for non-linear functional and more sophisticated relationships and lacks the flexibility to detect local patterns in the data. In contrast, non-parametric regression aims to directly model $E(y)$ without imposing restrictions on $h$. Mathematically speaking, $h(\vect x) = \sum_{r=1}^{\infty} \alpha_r \phi_r(\vect x) $ where $\phi_r$'s are known basis functions and $\alpha$'s are unknown regression parameters. Natural cubic splines, polynomial splines, B-splines, and radial basis functions are examples of basis functions. For a sufficiently large number $R$,
\begin{equation}     
h(\vect x) \approx \sum_{r=1}^{R} \alpha_r \phi_r(\vect x),
\end{equation}
and then one can proceed and estimate $\alpha$'s as usual. If $h$ is estimated by $\hat h$, then an estimate of $ E(y)$ at $\vect x_0$ is obtained by $g^{-1}(\hat h(\vect x_0 ))$. 
     
\begin{figure}
\centering
\includegraphics[width=\linewidth]{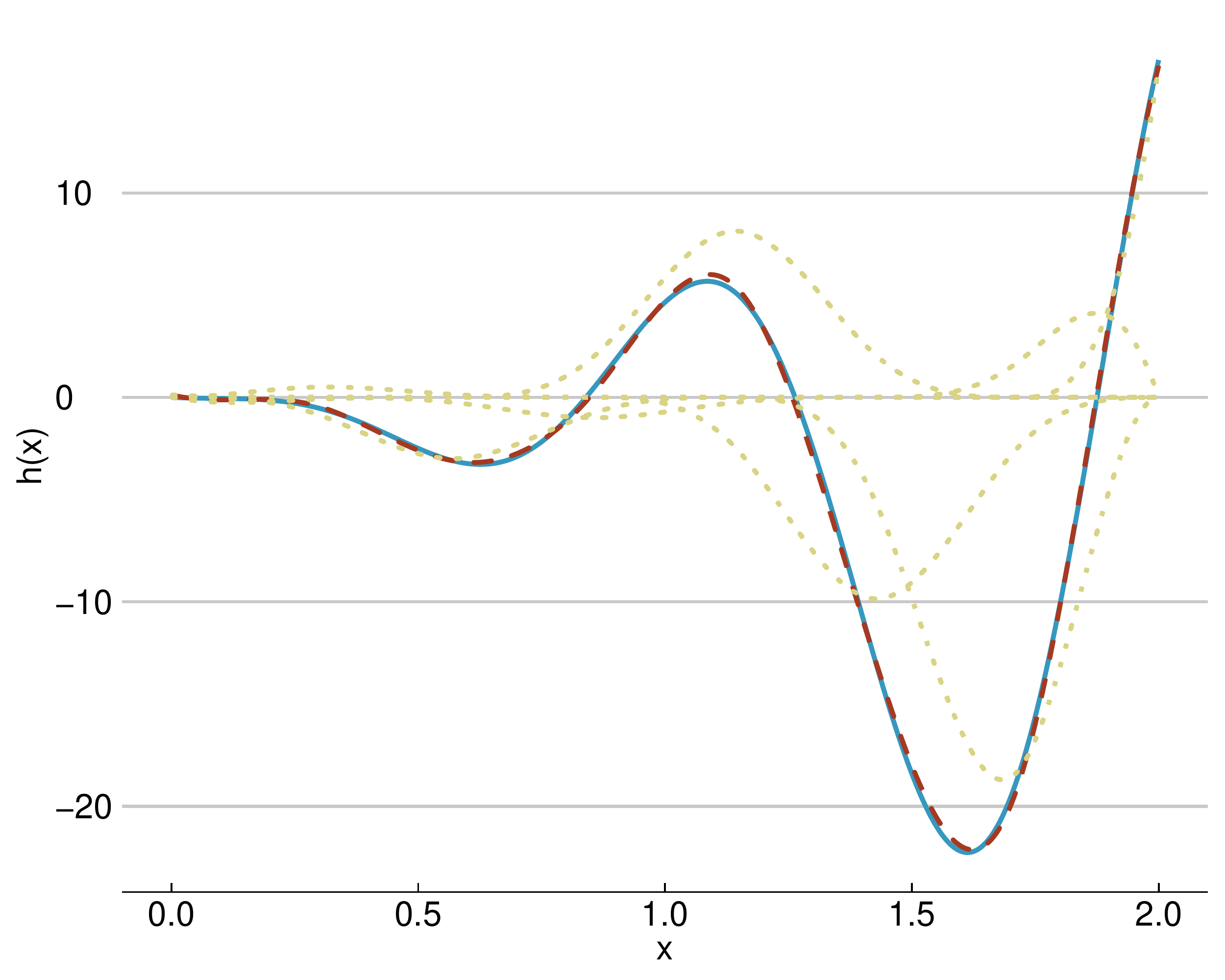}
\caption{Illustration of basis functions approach. The true regression function, $h(x)$, is represented by the solid blue curve. The basis functions are shown by the dotted lines, whereas the red dashed line gives the  reconstruction of $h(x)$, $\hat h(x)$.}
\label{fig:hx}
\end{figure}

For illustration, figure \ref{fig:hx} shows a reconstruction of a non-linear and non-monotonic regression function using a relatively small number of basis functions. The true function, $h(x) =   7x^2\cos6x-\exp x\tanh x$, is represented by the solid blue line curve. An estimate of $h(x)$, $\hat h(x)$, is obtained by taking a linear combination of $10$ polynomial splines represented by the dotted lines. Note that $\hat h(x)$ closely follows $h(x)$ despite using a limited number of splines. It is unlikely that using additional splines will result in a meaningful improvement.   

The parameter $R$ can be a fixed number determined manually by the user or by cross validation. Manual selection may lack flexibility and require multiple trials. Both approaches can result in ``wiggly" curves. A powerful approach is setting a relatively large number to $R$ and estimate $\alpha_r$'s using penalized  regression via maximum likelihood. This method will force insignificant effects  to be very close to 0 and thus it provides smooth representation of the data. Using MLE theory, conditional on $n$ being sufficiently large, one can conduct inferential statistics similar to GLMs as discussed before. Having said that, inferential statistics for non-parametric regression is more complicated. These details are out of the scope of this paper but they can be found in \citet{ruppert_wand_carroll_2003} and \citet{wood2006generalized}. A model that assumes $g(E(y)) = \vect x^T\vect \beta \label{eq:linearpre}$ is a special case of a model with  $g(E(y)) = h(\vect x)$. Thus, likelihood ratio tests can be used to test the linearity assumption.        
     
     Non-parametric regression does not impose functional forms on $h$ besides requiring $h$ to have continuous partial derivatives. Degree of complexity of $h$ can range from fairly simple to highly complicated. This allows to model sophisticated relationships that cannot be put in predetermined  forms,  and allows to reveal local patterns that cannot be detected using high order polynomials. However, this flexibility can be expensive. Complex patterns can be appropriately modelled if $n$ is sufficiently large. Even for a modest number of predictors, $h(\vect x)$ may require an inflated number $R$ for the approximation above to work properly. Reliable estimation of $h$ may not be obtained or be infeasible due to the curse of dimensionality. 
     
 As a remedy to this problem, \citet{Hast1990}  suggests to fit Generalized Additive Models (GAMs). GAMs assume that $h$ can be simplified into a linear combination of regression functions $h_j$'s. Specifically, $h(\vect x) = h_1(x_1) + h_2(x_2) + \hdots  +h_k(x_k)$ where each term $h_j$ is an unknown smooth regression function. For $f_{esc}$ data, $x_1 = M_{200}$, $x_2 = M_{\star}$ and so forth.   If each $h_j$  requires $m$ components then there is only $k\times m$ regression parameters to be considered. This allows to model a reasonably large number of predictors with a suitable sample size $n$. Furthermore, if one of the covariates, say $x_j$, is known to have linear relationship with $y$, then $h_j$ can be replaced by $x_j \beta_j$, achieving further reduction. This mixing between linear regression and non-parametric regression introduces what is called semi-parametric regression. 
     
   GAMs can be fitted using penalized least squares. Finally, if $h_j$ is estimated by $\hat h_j$, then $\mu_0$ is estimated by $\hat h(\vect x_0) = g^{-1}(\sum_{j=1}^{k} \hat h_j (x_{0j}))$. A $95\%$ confidence interval around $\mu_0$ can be obtained using penalized regression and MLE theory. Comprehensive details on fitting GAMs and conducting inferential statistics can be found in \citet{Hast1990}, \citet{christensen2001}, \citet{ruppert_wand_carroll_2003} and \citet{wood2006generalized}. See also \citet{Beck2017} for an application of GAMs to estimate the photometric redshift of galaxies, and \citet{Katrin06} for other non-parametric approaches named Gaussian Process. 
   
In what follows, we combine the concepts described in the previous sections to build the statistical model to probe  $f_{esc}$, which we call Hurdle Binomial-Beta Generalized Additive Model.    

\subsection{Hurdle Binomial-Beta Generalized Additive Model} 
\label{sec:Final}
This section applies the statistical framework described above to probe the relationship between $f_{esc}$ and the halo properties. The reasons behind the choice to employ this novel approach rather than a standard linear model can be simply summarized as follows.
The $f_{esc}$ shows an excess of zeros, hence requiring the need to use a two-part model (i.e. the hurdle binomial-beta model) for zeros and non-zero values.  The first component has a discrete nature that can take on two values $\{1,0\}$.   Such variables are naturally modelled using binomial models. The second component is a fractional property, for which a beta distribution rather than a linear model that depends on the normality assumption is better suited.  Finally, $f_{esc}$  has a non-linear relationship with the halo properties,  better described by a non-parametric model.

Based on our observations of $f_{esc}$ and its relationships with halo properties, we assume that $f_{esc}$ follows a hurdle model with logistic regression describing the probability of a galaxy with given properties to have $f_{esc} > 0$, and with beta regression with logistic link to evaluate the mean value of $f_{esc}$, $E(f_{esc})$, when $f_{esc} >0$. Furthermore, to account for the non-linear relationships between $f_{esc}$ and the halo properties, a generalized additive structure is assumed for the link functions.   

Specifically, the model assumes that the process $ \{f_{esc} > 0 \text{ vs }  f_{esc}=0 \}$,  denoted by $1_{f_{esc} > 0}$, 
has a binomial distribution such that
\begin{equation}
\displaystyle \log\frac{Pr\{f_{esc} > 0 \}}{Pr\{f_{esc} =0\}} = h_1(\vect x) = \sum_{j=1}^{7} h_{1j}(x_j),
\end{equation}
and the process $\{f_{esc} \text{ given that } f_{esc} > 0\}$, denoted by $f_{esc}|f_{esc} > 0$, 
has a beta distribution with mean $\mu$ in the form 
\begin{equation}
 \displaystyle \log \frac {\mu}{1-\mu} = h_2(\vect x) = \sum_{j=1}^{7} h_{2j}(x_j),  
\end{equation}
and a dispersion parameter $\phi$, where $x_1 = M_{\star}$, $x_2=M_{200}$, $x_3 = sSFR$, $x_4 = f_b$, $x_5 = \lambda$, $x_6=Q_{HI}$, $ x_7 = C$, and $h_{1j}\text{'s}$ and $h_{2j}\text{'s}$ are unknown regression functions. Following equation \ref{eq:Hurdle}, the average of $f_{esc}$ is 
\begin{eqnarray*}
E(f_{esc}) &=& p \times \mu \\&=& \left(1+\exp(-h_{1}(\vect x) \right)^{-1} \times \left(1+\exp(- h_{2}(\vect x)\right)^{-1}.
\end{eqnarray*}
\begin{figure*}
\centering
\includegraphics[width=\linewidth]{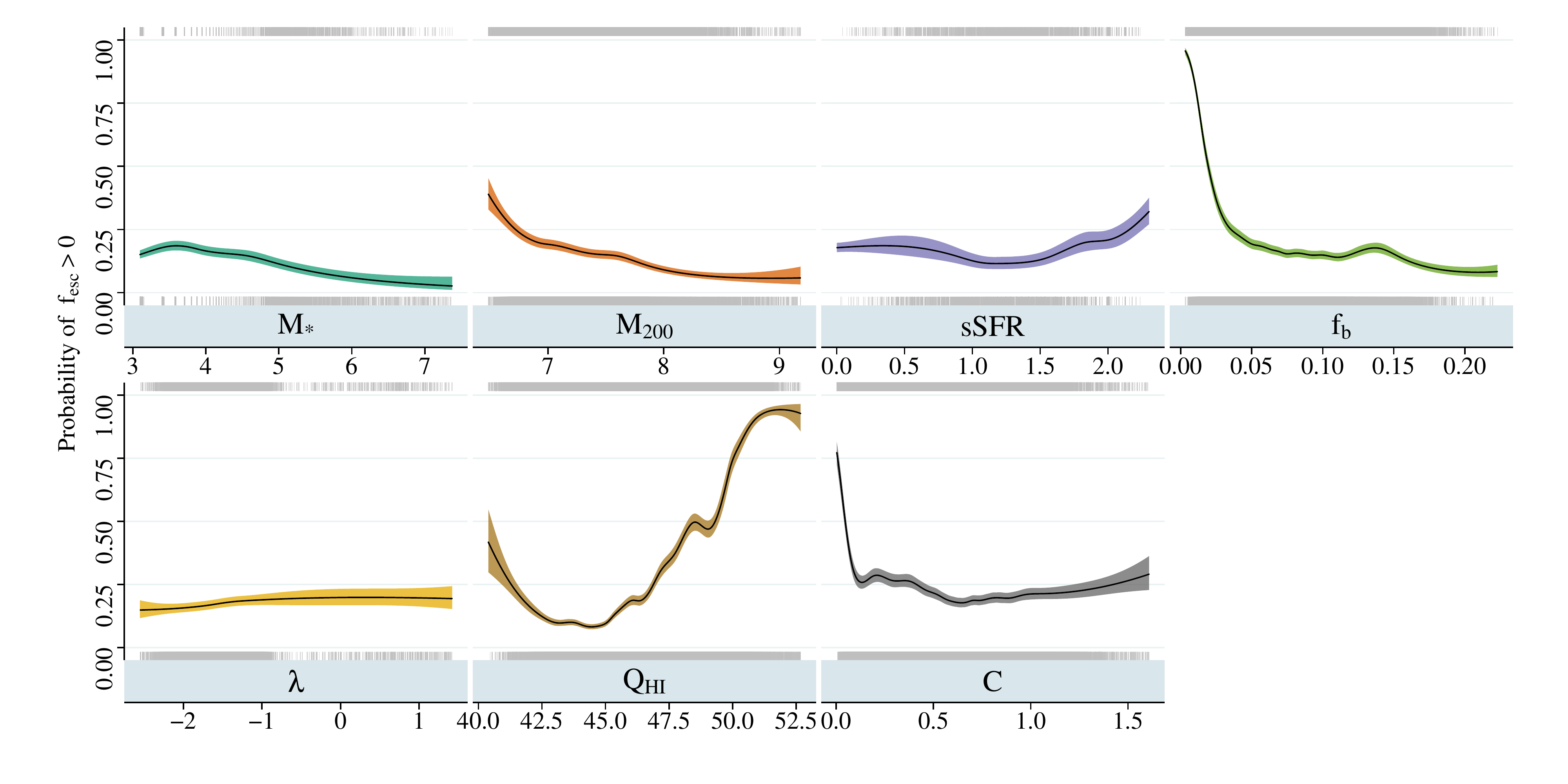}
\caption{Fitted probability curves for the seven galaxy properties indicated in the labels. In each panel, the black solid line represents the estimated probability of $f_{esc} > 0$ while varying only one galaxy property and holding other properties fixed at their median. The shaded areas depict $95\%$ confidence intervals. Data points ($1$ if $f_{esc} > 0  $ and $0$ otherwise) are laid out in the background.}
\label{fig:logit}
\end{figure*}

The estimation process is facilitated by  equation \ref{eq:hurdlelog},  which allows to separately fit a Binomial$-$GAM and a beta$-$GAM to $1_{f_{esc} > 0}$ and $f_{esc}|f_{esc} > 0$, respectively, as described in section \ref{sec:GAM}. Thus, point estimations and confidence intervals for each process features can be found in isolation of the other.

 Suppose that $h_1$ and $h_2$ are estimated by $\hat h_1$ and $\hat h_2$, and that $\vect x_0$ contains a specific combination of interest of galaxy properties, $\displaystyle \vect x_0^T = \left( M_{\star,0}, M_{200,0}, \hdots ,C_{0} \right)$. Then the estimated probability of $\displaystyle Pr\{f_{esc} > 0 \}$ at $\vect x_0$ is given by $\left(1+\exp(-\hat h_{1}(\vect x_0) \right)^{-1}$, the estimated average of $f_{esc}$ given that $f_{esc} > 0$ is $\displaystyle \left(1+\exp(- \hat h_{2}(\vect x_0)\right)^{-1}$, and the estimated average of $f_{esc}$  is 
           \begin{equation}
           \left(1+\exp(-\hat h_{1}(\vect x_0) \right)^{-1} \times \left(1+\exp(- \hat h_{2}(\vect x_0)\right)^{-1}.
           \label{eq:Avghur}
           \end{equation}
           The estimated standard error of this quantity can be found using equation \ref{eq:SE} and an approximate confidence interval of $E(f_{esc})$ follows. This standard error reflects the uncertainty from estimating $E(f_{esc})$ with equation \ref{eq:Avghur}. 
           
     According to equation \ref{eq:meanvarhur}, the variance of $f_{esc}$ is given by 
 \begin{equation}    
\displaystyle p\left[\frac{\mu (1-\mu)} {1+\phi} + (1-p) \mu^2\right],   
 \end{equation} 
and measures the uncertainty of $f_{esc}$ around $E(f_{esc})$. It is estimated by replacing $p$, $\mu$ and $\phi$ with their estimates $\hat p$, $\hat \mu$ and $\hat \phi$.   

The odds of $f_{esc} >0$ at two points $\vect x_1$ and $\vect x_2$ are $\displaystyle \exp(h_1(\vect x_1))$ and $\exp(h_1(\vect x_2))$. The odds ratio between the two points is given by $\displaystyle \exp( h_1(\vect x_1) - h_1(\vect x_2))$, and it is evaluated by replacing $h_1$ with its estimate $\hat h_1$. If it is greater than $1$, it indicates that the odds of a galaxy with halo properties $\vect x_1$ to produce enough photons capable to escape the halo  is greater than for a galaxy with halo properties $\vect x_2$. A confidence interval of $h_1(\vect x_1) - h_1(\vect x_2)$ can be  obtained from the Binomial-GAM fit. Exponentiating this interval gives a confidence interval for the odds ratio. For example, if the probabilities of $f_{esc} >0$ at $\vect x_1$ and $\vect x_2$ are $0.015$ and $0.012$ respectively, the difference in terms of probabilities is $0.003$ and the odds ratio is $1.25$, meaning that moving from $\vect x_2$ to $\vect x_1$ increases the odds of observing a positive value of $f_{esc}$ by $25\%$. 

If $\vect x_1$ is identical to $\vect x_2$ except at the $j$th component, then the odds ratio is $\displaystyle \exp( h_{1j}(x_{1j}) - h_{1j}(x_{2j}))$. For comparison purposes, the non-$j$th components of $\vect x_1$ and $\vect x_2$ are usually set to their medians. Notice that a full specification of both vectors is needed to compare $\vect x_1$ and $\vect x_2$ in terms of probabilities, whereas the odds ratio interpretation holds regardless of the settings of the non-$j$th components, as long as they are held fixed. Although probabilities provide a more straightforward interpretation, both quantities are relevant to understand the importance of the covariates to the process $1_{f_{esc} >0}$.       

This model addresses to a large extent the five limitations of Model \ref{eq:modelone} that was discussed in Section \ref{sec:LM}. Although we do not imply that this model is fully correct 
\footnote{After all, \textit{Essentially all models are wrong but some are useful} \citep{Box87}.},
it has a degree of flexibility that can deal with many aspects of $f_{esc}$ and its complex relationships with halo properties without fully compromising the physical interpretation of the results. 
\begin{figure*}
\centering
\includegraphics[width=\linewidth]{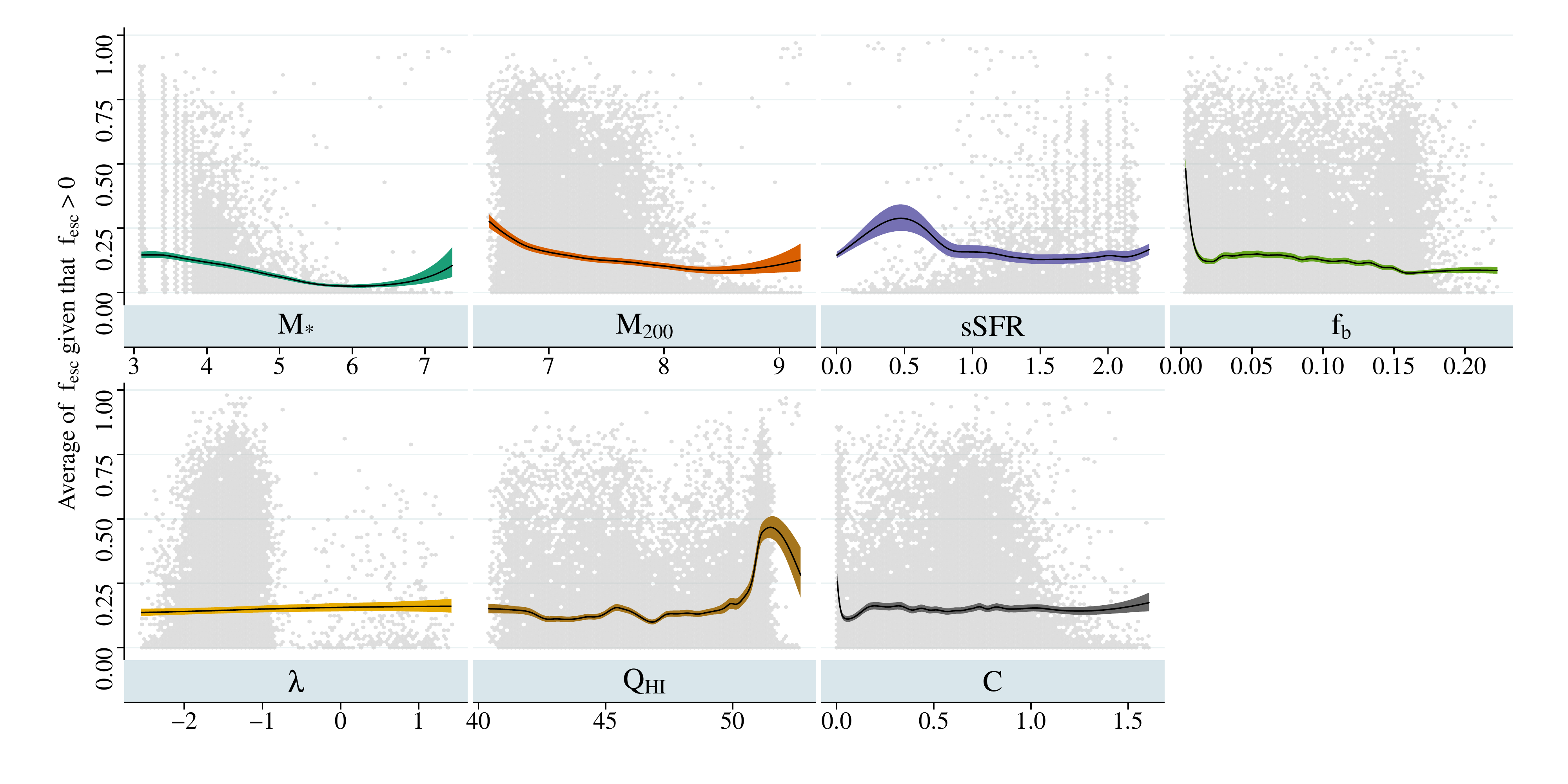}
\caption{
Estimated mean curves of $f_{esc}$ when $f_{esc} >0$ for the seven galaxy properties indicated in the labels. In each panel, the black solid line represents the estimated mean while varying only one galaxy property and holding other properties fixed at their median. The shaded blue areas depict $95\%$ confidence intervals. Data points are laid out in the background.
}
\label{fig:beta}
\end{figure*}

\section{Modelling Escape Fraction: Results}
\label{sec:Results} 

This section summarizes the results obtained by fitting to the $f_{esc}$ dataset the Hurdle Binomial-Beta Generalized Additive Model introduced in Section \ref{sec:Final}. 

We first fit a Binomial-GAM with a logistic link to $1_{f_{esc} > 0 }$ responses as described in \ref{sec:Final}, and we denote this by model $M_1$. Approximately $32.6\%$ of the  variability of $1_{f_{esc} >0 }$ is accounted for by model $M_1$, a measure that has been adjusted to take into account the number of regression parameters in the model. A standard diagnostic for binomial regression is the so-called area under the curve \citep[AUC; see e.g.][]{deSouza2014}, which  can be used to measure the discriminatory ability of model $M_1$ to distinguish between galaxies for which ionizing radiation is able to escape into the IGM and those for which this does not happen. AUC values range between $0$ and $1$, and an AUC of $0.5$ corresponds to random guessing, while  AUC = $1$ indicates perfect discriminatory power. The AUC of $M_1$ is $0.82$, indicating a fairly high discriminatory power. 

For comparison, a logistic regression model, $S_1$, that assumes linearity via
\begin{equation}
 \log\frac{Pr\{f_{esc} > 0 \}}{Pr\{f_{esc} =0\}} = M_{\star} + M_{200} + \hdots + C, 
\end{equation}
is fitted and tested against $M_1$ using LRT as explained in \ref{sec:GAM}. The observed value of this test statistic is $11360$. As compared against a $\chi^2$-distribution (details can be found in \citealt{wood2006generalized}), the test rejects $S_1$ with a massive evidence in favour of $M_1$. Furthermore, the BIC is $86,208$ for $S_1$ and $75,628$ for $M_1$. Although such comparisons are not needed since it is apparent that $S_1$ cannot adequately describe such complex relationships, we show them to emphasize and encourage the use of GAMs whenever is needed. A BIC of $75,690$ has been obtained by fitting model $M_1$ with a probit link rather than a logistic link, indicating that the latter is slightly more plausible.  

Each panel of figure \ref{fig:logit} corresponds to the estimated probability with $95\%$ confidence intervals that $f_{esc} > 0$ when one galaxy property is varied while holding all the others fixed at their medians. As an example, for the $Q_{\sc HI}$ panel, the estimated probability that a galaxy with halo properties $\vect x_1 \equiv  (M_{\star} \equiv 3.39, M_{200} \equiv 7.15, sSFR\equiv0, f_b \equiv 0.06,\lambda \equiv -1.47, Q_{\sc HI} \equiv 45, C \equiv 0.65)$ 
will have non zero escape fraction is $0.095$ with $(0.084, 0.107)$ as a $95\%$ confidence interval. Whereas the probability at another point $\vect x_2$ which is identical to $\vect x_1$ except $Q_{\sc HI}$ now equals $50$ rather than 45 is $0.737$ with $(0.693,0.777)$ as a $95\%$ confidence interval. The odds that $f_{esc} > 0 $ increase by $24$ folds when moving from $\vect x_1$ to $\vect x_2$. This holds regardless of the other galaxy properties being set to their medians or to any other value, as long as they were held fixed. 
 On the other hand, computing probabilities or the difference in probabilities requires full specification of all covariates. The probability of $f_{esc} > 0$ at $\vect x_1^\prime$ and $\vect x_2^\prime$ where $Q_{\sc HI}=45$ and $Q_{\sc HI}=50$ and the other covariates are held fixed at their first quartile rather than at the median is $0.187$ and $0.845$, respectively. The probabilities and their difference have changed but the odds ratio stayed the same.

\begin{figure*}
\centering
\includegraphics[width=\linewidth]{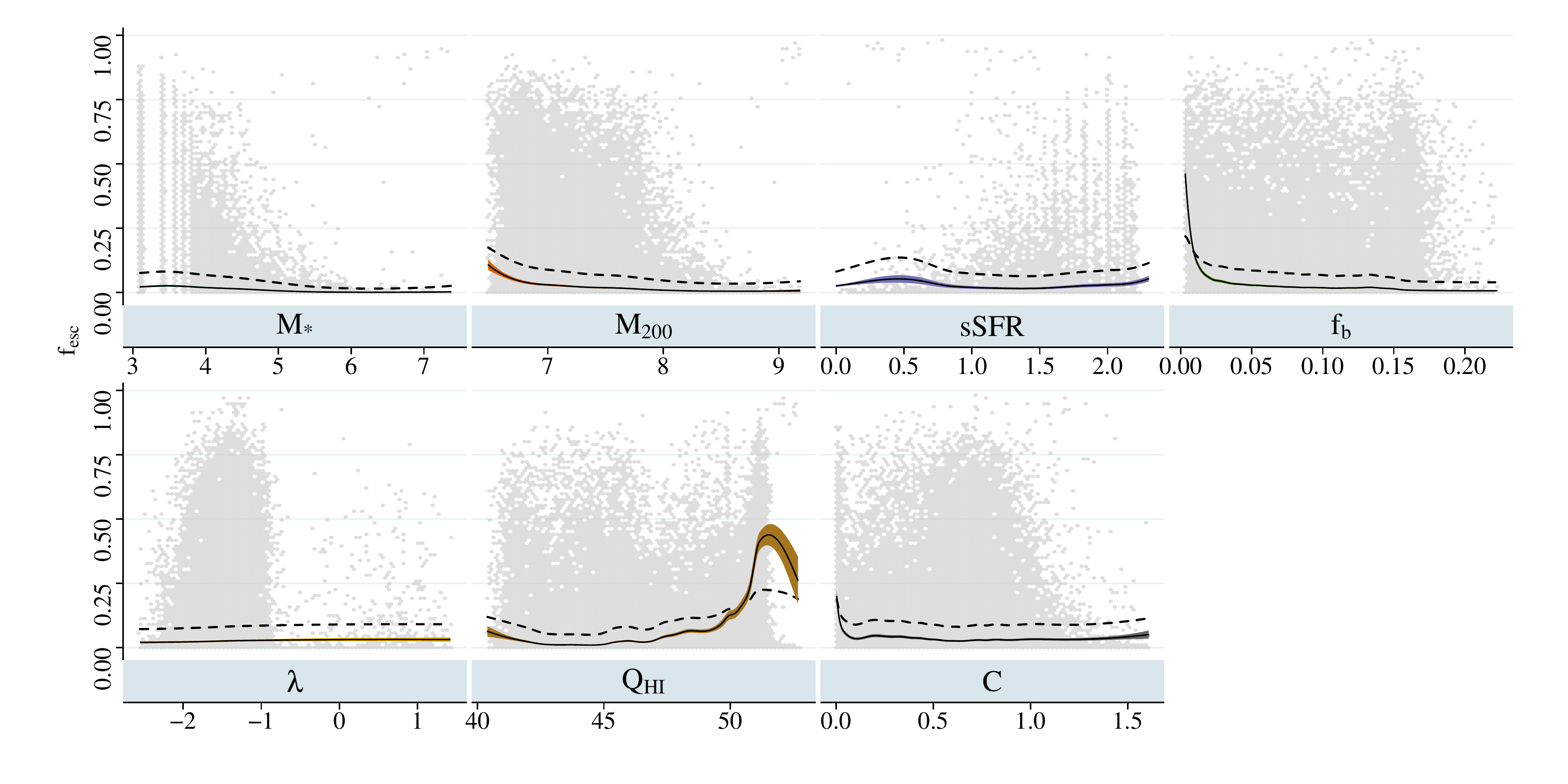}
\caption{ 
Estimated mean curves of $f_{esc}$ for the seven galaxy properties as indicated in the labels. In each panel, the black solid  line represents the estimated means while varying only one galaxy property and holding other properties fixed at their median. The shaded areas depict $95\%$ confidence intervals. The dashed line is the estimated standard deviation of $f_{esc}$. Data points are laid out in the background.} 
\label{fig:hurdle}
\end{figure*}

To investigate the predictive ability of the fitted model, the galaxies are divided into two subsets. $M_1$ is fitted to the first subset, which makes up $75\%$ of the dataset, independently of the second subset. Then $M_1$ is used to predict the responses in the second subset. 
An AUC of 0.815 has been obtained, falling short only by $0.005$ as compared to the fitted model using the whole data. The procedure has been repeated multiple times (using different splits each time) yielding very similar results and confirming the predictive ability of $M_1$.  

Next, we fit a Beta-GAM to $f_{esc}|f_{esc} > 0$ responses as described in \ref{sec:Final}, and we denote this by model $M_2$.
The sample size is 29,769 after removing the responses that are exactly zero. 
Figure \ref{fig:beta} represents the estimated average of $f_{esc}|f_{esc} > 0$ along with $95\%$ confidence intervals when varying a predictor and holding all other predictors fixed at their medians. 

The statistical properties of $f_{esc}$ as a whole can be obtained by combining information from $M_1$ and $M_2$. Figure \ref{fig:hurdle} represents the estimated average of $f_{esc}$ along with $95\%$ confidence intervals according to equations \ref{eq:Avghur} and \ref{eq:SE}. 
Basically, the curves in figure \ref{fig:hurdle} are a multiplication of the corresponding curves in figures \ref{fig:logit} and \ref{fig:beta}. The result of this operation is that, while the shape of the curves remains very similar, due to presence of zeros the curves in figure \ref{fig:hurdle} are closer to 0 that those in figure \ref{fig:beta}. 
Also, the width of the confidence intervals has substantially decreased. The natural variability of $f_{esc}$ is reflected by the estimated standard deviation (dashed lines), which, in most cases, is greater than the mean. 

\begin{figure*}
\centering
\includegraphics[width=\linewidth]{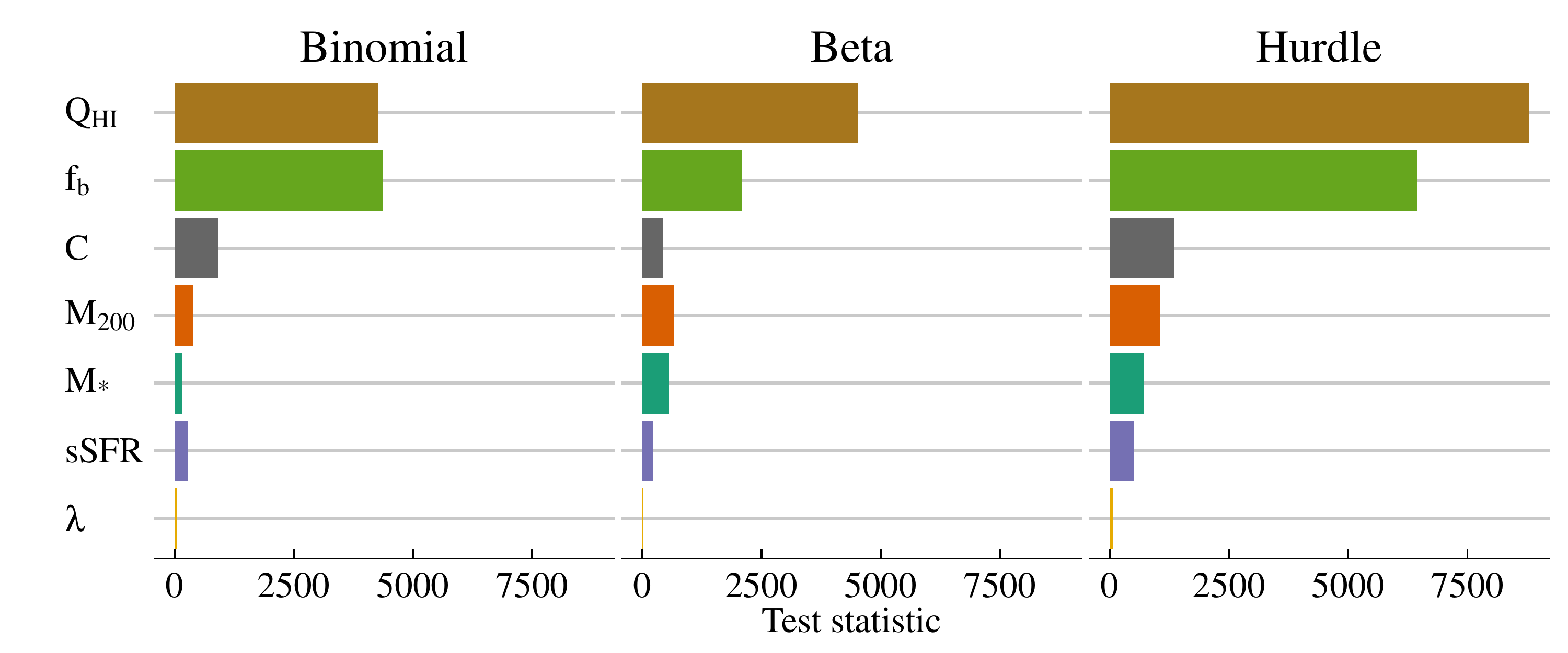}
\caption{ 
Galaxy properties ordered according to the value of their test statistic minus the test degrees of freedom for each part of the hurdle model. Properties closer to the top on the rightmost panel have higher influence on $f_{esc}$.} 
\label{fig:test_h}
\end{figure*}

Next we examine the importance of galaxy properties. Broadly speaking, the shape and steepness of the curves in the previous figures can serve as indicators of the predictors relative influence on $f_{esc}$. A more formal approach to evaluate the importance of the various properties is to use the Wald test statistic, $T_{kj}$, for assessing the significance of the $j$th predictor in model $M_k$ for $k=1,2$ after taking the other predictors into account. According to section  $\ref{sec:hurdle}$, $T_{1j} + T_{2j}$ is the test statistic for assessing the significance of the $j$th predictor for both processes altogether. Specifically, it is the test statistic for testing the hypothesis that the $j$th predictor has no influence on the probabilistic threshold above which the photons are capable to escape the galaxy and on the mean value of $f_{esc}$ if photons escape. Galaxy properties have been ordered according to the value of this test statistic minus the test degrees of freedom. This ordering is reported in Figure \ref{fig:test_h} for each part of the model and quantitatively confirms what has been more qualitatively illustrated in the previous figures.

We find that $Q_{\sc HI}$ and $f_b$ are by far the most influential properties. Conversely, $C$, $M_{200}$, $M_{\star}$ and $sSFR$  play less important roles, while $\lambda$ seems to have a negligible influence. The test suggests also that while $C$ has more influence in defining  the probability of photons escaping the galaxy (i.e. defining whether $f_{esc}$ is 0 or not), $M_{200}$ and $M_{\star}$ are more important in the regime in which $f_{esc} > 0$. With $f_{esc}$  increasing with decreasing halo mass below $M_{200} \lesssim 10^9 M_{\odot}$, stressing the role of the smallest galaxies as sources of ionizing radiation \citep[e.g.][]{Xu2016}.

The dominance of $Q_{\sc HI}$ is due to its direct relation with the ionization rate, and thus the ability to create paths free from neutral hydrogen. 
While its importance in comparison to e.g. $M_{\star}$ and $M_{200}$ might be counterintuitive, it should be noted that $Q_{\sc HI}$ is mostly  dependent on young stellar population, unlike $M_{\star}$ and $M_{200}$ that correlate with all stars, young and old. 
While there is a correlation between SFR and $M_{\star}$, this is only an average effect, and it does not take into account fluctuations in the SFR for a given $M_{\star}$. Hence any correlation with $M_{\star}$  or $M_{200}$ is not as strong as that directly with $Q_{\sc HI}$.
As expected, galaxies with a high $Q_{\sc HI}$ have larger $f_{esc}$ as a consequence of both a higher production of ionizing photons and, depending on their star formation history, a stronger supernova feedback. A relation to feedback was observed also in  \citet{Kimm2014}, who found a time delay   between the peak of star formation and that of $f_{esc}$, due to the time required for subsequent destruction of the star-forming cloud by supernova feedback. A strong dependence on stellar feedback was confirmed also by \citet{Ma2015}: although the intrinsic budget of ionizing photons is dominated by young stellar populations as expected in standard population models, the majority of escaping photons derives from intermediate age stars located in clouds which have been cleared by feedback effects.

The second most influential property is $f_b$ through its relation with the (column) density of gas, i.e. higher $f_b$ shows a distribution that has higher column densities around photon sources. More specifically, smaller haloes (which typically have also smaller sSFR and $M_{\star}$), as well as those with a low baryon fraction, have higher escape fractions because of the lower gas column density encountered by the photons on their way to the intergalactic medium.
While previous works \citep[see e.g.][]{CiardiFerrara2005} determined a strong dependence of the escape fraction on the gas distribution ($f_{esc}$ increasing with the clumping factor), here we find only a mild dependence, possibly because of the relatively small values of $C$ obtained due to the limited resolution of the interstellar medium.

It should be underlined that these results largely depend on the predictors being weakly to moderately correlated. Therefore, due to the complex correlation structure of the covariates existing in this dataset, they should be taken with caution. Notice that a naive univariate analysis using correlations suggests smaller effects for $f_b$ and $Q_{\sc HI}$ and a larger impact for $sSFR$ (see Fig. \ref{fig:all}). In particular, the influence of $sSFR$, which reflects the balance between dense gas (hence low escape fraction) promoting star formation and high number of ionizing photons (hence high escape fraction), becomes far smaller after taking the other predictors into account and after accounting for non-linearities in a regression context.

\section{Discussion and conclusions}
\label{sec:results_discussion}

The epoch of reionization represents a milestone in the history of the Universe, and a result from very complex interactions between photons yield from ionizing sources and their surrounding environment. A key proxy for the ionizing power  of a given  source (e.g. first generation of stars, quasars, etc.) is the escape fraction, i.e. the fraction of photons effectively capable to reach the intergalactic medium. This is a fractional, hence non-Gaussian,  physical property that relates non-linearly to the properties of its host galaxy.  

During the course  of cosmic evolution,  scaling  relations between the escape fraction and galaxy properties emerge once a physical threshold is transposed, i.e. a halo needs to reach a minimum mass capable to form the first stars and subsequently produce ionizing photons. In statistical parlance, such relations can be probed by the so-called hurdle models \citep[e.g.][]{Hilbe2017}.

As a case in point, we have applied the hurdle model to describe the dependence of the escape fraction on several halo properties, and have introduced a statistical criterion to rank such properties according to their influence on the escape fraction, accounting for the non-linear nature of the relation between the various quantities.   

This analysis shows that the production rate of ionizing photons, $Q_{\sc HI}$, and baryonic fraction, $f_b$, are the most influential galaxy properties, emphasising that the interplay between star formation for the production of ionizing photons and supernova feedback for clearing away dense gas is the process that determines the escape fraction. A naive univariate analysis suggests smaller effects of $f_b$ and $Q_{\sc HI}$, but a much larger impact of the specific star formation rate. 

From a methodological viewpoint, because of the statistical model ability to \textit{interpolate} across the multidimensional space of finite sampled simulations, the approach  acts as an emulator to estimate $f_{esc}$ in between sampled simulated points.  Thus, it  provides the means to fast generate samples for specific values of galaxy properties in situations where the computational cost of a full simulation is too expensive. 

Finally, we showed how linear models, widely used in astronomy,  can be readily extended to cover complicated datasets. We started presenting its technical material and clearly stating the assumptions that make the use of linear models appropriate. Linear models are the second to none if these assumptions are not violated. However, in astronomy, these assumptions are frequently not satisfied, and more flexible models have been developed to  defeat these limitations \citep{deSouza2015AeC,Elliott2015,deSouza2015MNRAS.453,deSouza2016MNRAS}.  Fitting procedures and conducting statistical inferences have been reviewed not only for linear models but also for generalized linear models, binomial regression, beta regression and generalized additive models. Each of the previous term can find plenty of applications in Astronomy.  These models can be coherently combined using hurdle models to probe observations resulting from a mixture of underlying physical processes.

We therefore advocate for the use of  hurdle,  GAMs, and its variants,  given  its potential to become a valuable statistical tool  for Astronomers due to its richness and ability to adapt to complexities that are usually encountered in the field.

\section*{Acknowledgements}

The authors thank the anonymous reviewer, and Eric Feigelson for their insightful comments. 
RSS acknowledges the support from NASA under the Astrophysics Theory 
Program Grant 14-ATP14-0007. JPP acknowledges support from the European Research Council under the European Community's Seventh Framework Programme (FP7/2007-2013) via the ERC Advanced Grant "STARLIGHT: Formation of the First Stars" (project number 339177).
We thank the  Cosmostatistics Initiative\footnote{\href{https://cosmostatistics-initiative.org}{https://cosmostatistics-initiative.org}} (COIN) -  where this interdisciplinary research team were triggered. COIN is a non-profit organization whose aim is to nourish the synergy between astrophysics, cosmology, statistics and machine learning communities.

\bibliographystyle{mnras}
\bibliography{ref}
\end{document}